# Vacuum Moisture Swing Direct Air Capture: A Low-Thermal, Water-Managed Pathway for Scalable $CO_2$ Removal

**Stephano Sinyangwe[1], Sierra Binney[1], Gray Becker[1], Peter Schulze[2], Jennifer L. Wade[1*]**

[1] The Steve Sanghi College of Engineering, Mechanical Engineering, Northern Arizona University, Flagstaff, Arizona, 86011, USA

[2] Scientific Chemical Process Engineering Consultancy (SChPrEngCo), 39218 Magdeburg, Germany

## Abstract

Direct Air Capture (DAC) remains one of the most energy-intensive carbon removal approaches, with most systems relying on high-temperature amine or metal oxide regeneration. Here we present the first comprehensive evaluation of a feasible, low-temperature DAC process that exploits the moisture-swing mechanism to reversibly capture and release $CO_2$ using commercial quaternary ammonium exchange resins (IRA 900). The proposed vacuum moisture swing (VMS) sorption technology introduces a five-step process that replaces high-temperature $CO_2$ regeneration with a low-temperature water vapor stripping step driven by vacuum evaporation. Using experimentally measured water and $CO_2$ sorption kinetics, the cyclic model was optimized across a range of air relative humidities (20-80%) and kinetic regimes (0.5x-2.0x baseline).

Results show that optimized VMS operation in 20% RH air achieves $CO_2$ productivities of 0.2-0.6 kg $\mathrm{CO_2} \cdot \mathrm{kg_s^{-1}} \cdot \mathrm{d^{-1}}$, comparable to or exceeding those of high-temperature amine sorbent systems, yet without the addition of external heat. The required electrical energy to drive the gas / vapor flows and compress $CO_2$ to 0.1MPa spans a broader range of 1-15 $\mathrm{MJ} \cdot \mathrm{kgCO_2}^{-1}$, due to a strong inflection of energetic loads that is largely governed by the vapor-flow rates in the vapor-stripping step. At a representative productivity of 0.5 kg $\mathrm{CO_2} \cdot \mathrm{kg_s^{-1}} \cdot \mathrm{d^{-1}}$, the energetic load is limited to ~2.5 $\mathrm{MJ} \cdot \mathrm{kgCO_2}^{-1}$, surpassing the productivity-energetic trade-off of nearly all DAC systems, but incurs water losses of 1.4-3.5 $\mathrm{kgH_2O} \cdot \mathrm{kgCO_2}^{-1}$ through evaporation, similar to liquid-adsorption DAC systems. Water loss scales linearly with productivity and decreases at higher humidity and faster kinetic conditions. Importantly, the VMS process manages water through vacuum-driven evaporation and condensation, enabling the use of non-fresh or saline water streams and thereby reducing the environmental footprint of DAC deployment.

This work introduces the first experimentally validated low-temperature DAC pathway that integrates realistic $CO_2$ and $H_2O$ transport behavior with built-in water management. By achieving strong performance under variable humidity and without external heat, the VMS framework establishes a new design space for scalable, resource-efficient DAC and sets the stage for techno-economic and lifecycle evaluation of the next generation DAC systems.

**Keywords:** Direct Air Capture, $CO_2$ separation, sorption, vacuum moisture swing

# 1.0 Introduction

The urgent need for carbon dioxide $CO_2$ removal from the atmosphere has gained global attention as an essential component in achieving carbon neutrality by 2050.[1–3] The increasing frequency and severity of climate change impacts underscore the necessity for multiple innovative and scalable negative emissions technologies (NETs).[2,4–6] While conventional emission reduction efforts such as renewable energy, nuclear power and carbon capture and storage ( CSS) adoption have advanced, these measures alone are insufficient to address the historical and ongoing accumulation of $CO_2$ in the atmosphere.[7] Direct air capture (DAC) – the separation of $CO_2$ directly from the atmosphere – with permanent and safe geologic sequestration is emerging as a potential strategy to bridge this gap due to its capability to extract $CO_2$ at dilute concentrations (~0.04%).[8–10] Additionally, DAC offers high negative emission rates and flexible siting compared to other NETs like afforestation/reforestation or bioenergy with CSS (BECSS).[11,12] Despite these advantages, this technical approach still faces significant economic and energy barriers, necessitating innovations to ensure its feasibility at a commercial scale.[13,14]

The leading DAC technologies under commercial development include aqueous scrubbing with either alkaline hydroxides, amino acid solvents or solid sorption processes using supported amines or alkaline earth metal oxides.[13,15,16] Both aqueous and sorbent-based processes have distinct advantages and disadvantages. Solvent-based separations, like potassium hydroxide (KOH), can operate continuously allowing for high productivity but yet face limitations as far as process complexity (and in turn cost) due to high temperature regeneration.[16] In this method, dilute $CO_2$ reacts with the solvents in an air contactor to form dissolved carbonates ($CO_3^{2-}$). The resulting $CO_2$ loaded solvent is moved into a pellet reactor filled with a calcium hydroxide ($Ca(OH)_2$) solution, in which the trapped $CO_2$ as $CO_3^{2-}$ is precipitated by reaction with $Ca^{2+}$ forming calcite, $CaCO_3$. To release concentrated $CO_2$ the $CaCO_3$ must be calcined at high temperatures ( ~900°C) accounting for at least 60% of the total energy demand of this separation.[16,17] Furthermore, the process incurs high water loss rates in the air contactor as water from the solvent evaporates in air. The evaporated water loss varies with ambient conditions and solvent concentration, ranging anywhere from 3-9 $kg_{H_2O} \cdot kg^{-1}_{CO_2}$.[16] In the case of dissolved amino acids, their organic nature provides fast $CO_2$ reaction rates, low volatilities, and low toxicities, with the low volatility being particularly advantageous as it reduces water loss in the air contactor.[18,19] Custelcean *et al.*[13] reported that regenerating amino acids such as glycine or sarcosine with solid iminoguanidine, rather than alkaline earth metal oxides, allows for lower temperature regeneration (80°C-120°C). Despite these advantages, amino acid solvents are relatively energy-intensive and can degrade over prolonged use.

In contrast, solid sorption systems utilize amine-supported materials like mesoporous silicas, polymers, and metal organic frameworks (MOFs), which offer selective $CO_2$ binding and high surface areas. The supported amines, for instance, can be chemically grafted onto porous substrates, allowing for high $CO_2$ selectivity and rapid sorption kinetics under ambient conditions. These sorbents typically require less energy for regeneration compared to solvent methods, making them a potentially more energy-efficient option.[17] However, thermal oxidative degradation of amines during cyclic operation reduces their long-term stability and operational lifespans to a few years.[20]

A key challenge in DAC is developing chemisorbing materials with fast kinetics, adequate working capacity and durable stability that enables effective performance for over $10^4$ cycles, criteria essential for reducing both environmental impact and operating costs.[21] Advances in sorbent design must be paired with process innovations that fully leverage the strengths of these materials. Recent process-optimized studies of amine-supported sorbents, have focused on increasing sorption capacity, improving sorption kinetics and enhancing thermal oxidative stability.[22–25] Together, these developments aim to reduce amine degradation while optimizing $CO_2$ capture productivity. High-capacity metal oxide sorbents, such as magnesium (24.8 mmol/g) and calcium oxides (17 mmol/g), offer an alternative route, offering sorbent stability and durability.[26,27] Nevertheless, developing cost-effective sorbents that balance long term stability with high $CO_2$ uptake remains critical for practical deployment.[28]

Among the various regeneration methods for sorbent-based DAC, Temperature Vacuum Swing Adsorption (TVSA) and steam-assisted TVSA (s-TVSA) have emerged as the most widely adopted sorbent-based DAC configurations.[29–33] Both rely on temperature and pressure swings to release $CO_2$ at high purity and reduce the oxygen that exacerbates thermal oxidation. The steam-assisted process further enhances $CO_2$ transport during regeneration to increase productivity, albeit with added thermal duty for steam generation and the capital expenditure of a steam plant. Although effective, these regeneration pathways require substantial thermal and electrical energy (approximately 5-7 $\mathrm{MJ \cdot kgCO_2}^{-1}$), which accounts for a significant portion of the overall energy demand in the DAC process.[13] Solely relying on pressure-based techniques like Pressure Swing Adsorption (PSA), Pressure Vacuum Swing Adsorption (PVSA) and Vacuum Swing Adsorption (VSA) are generally unsuitable for DAC because the required air compression or deep vacuum (< 40Pa) imposes prohibitive energetic and economic penalties.[34] An interesting exception arises when cryogenic fluids are available, enabling highly selective $CO_2$ uptake on physical sorbents like zeolites cooled by decompressed liquefied natural gas; however, such opportunities remain limited and reliant on an ongoing fossil-based infrastructure.[35]

These limitations have motivated a growing interest in alternative regeneration strategies that avoid external heating altogether. Moisture-swing (MS) sorbents, which reverse their $CO_2$ affinity through changes in water activity rather than temperature, offer a fundamentally different pathway that can circumvent the high thermal-energy burdens of TVSA systems.[36–39] By coupling water-mediated binding chemistry with pressure-driven evaporation, these sorbents enable low-temperature regeneration and open new opportunities for energy-efficient DAC process design.

MS sorbents are typically anion exchange resins with positively charged quaternary ammonium groups. They operate by binding ambient $CO_2$ as bicarbonate when dry and releasing $CO_2$ upon exposure to moisture. A summary of the net mechanism can be seen in eqn (1)

$$2HCO_3^{-}.mH_2O + (n\text{-}m\text{-}1)H_2O \rightleftharpoons CO_3^{2-}.nH_2O + CO_2 \quad (1)$$

where the stoichiometric coefficient of water, $(n - m - 1) > 0$, satisfies mass action requirements and Le Chatlier's principle. The hydration shells surrounding $HCO_3^-$ and $CO_3^{2-}$ dictate the direction of the equilibrium. Under humid conditions, additional water stabilizes carbonate ions driving $CO_2$ release; under dry conditions, dehydration destabilizes carbonate, shifting the equilibrium back toward bicarbonate and hydroxide, whereby the hydroxide readily binds with the low concentrations of $CO_2$ in air.[39–41]

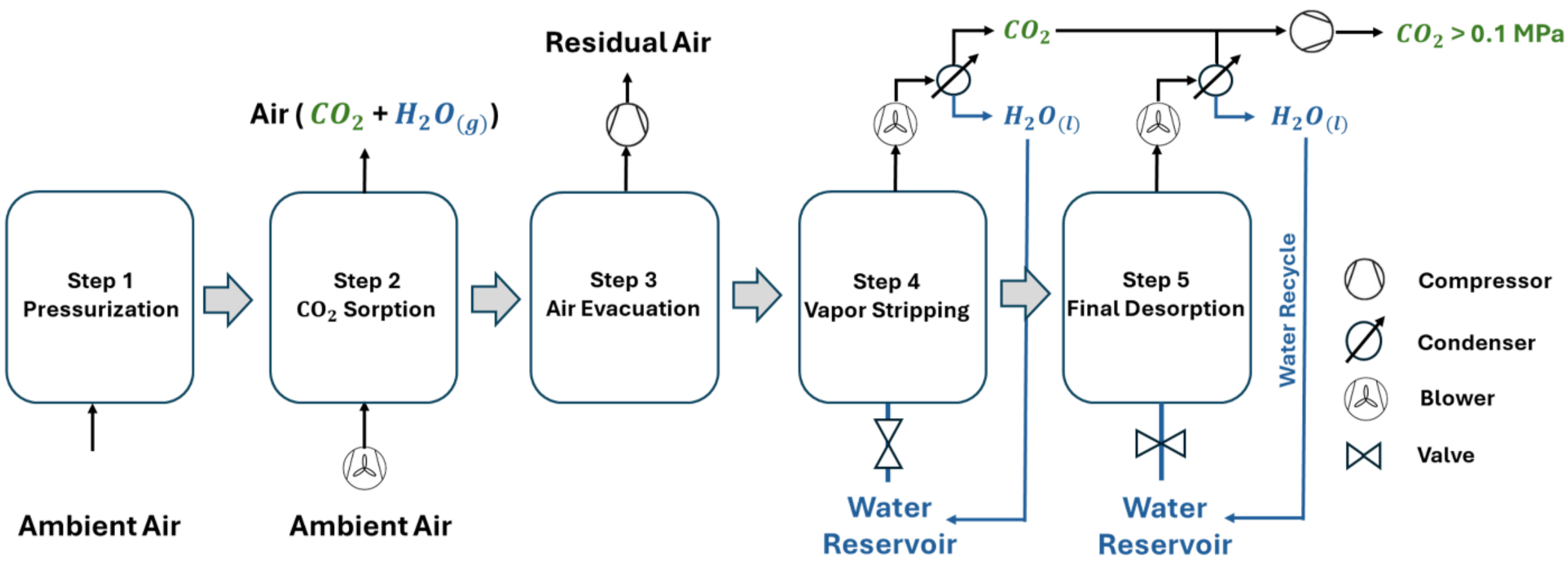

**Fig. 1** Vacuum Moisture Swing sorption (VMS) for dilute $CO_2$ separation, consisting of five key steps: (1) Pressurization, (2) $CO_2$ Sorption, (3) Air Evacuation, (4) Vapor Stripping, and (5) Final Desorption

This moisture-mediated mechanism for reversible $CO_2$ sorption minimizes the need for thermal heating of the sorbent, reducing dependence on fossil fuel derived energy and representing a low energy, fossil free $CO_2$ removal process. To leverage these advantages, the Vacuum Moisture Swing (VMS) process is proposed as a regeneration strategy. This technology combines the MS mechanism with vacuum application to evaporate water and evacuate the desorbed $CO_2$. The proposed VMS process comprises five steps (Fig. 1), beginning with the column initialized in its desorbed state. 1) **Pressurization**: air is introduced through the inlet to the sorbent bed to raise the column pressure back to ambient. 2) **$CO_2$ Sorption**: blowers push air past the sorbent bed, allowing $CO_2$ to bind to the sorbent while simultaneously evaporating $H_2O$ out of the bed from the wet $CO_2$ desorbed sorbent. 3) **Air Evacuation**: with the inlet closed, vacuum is applied at the outlet to remove residual air to increase $CO_2$ purity. 4) **Vapor Stripping**: both inlet and outlet are open under vacuum conditions, promoting continuous water evaporation through the column, while stripping $CO_2$ away from the sorbent bed. 5) **Final Desorption**: the water inlet is closed, and residual $CO_2$ and $H_2O$ are released from the outlet under vacuum, while also reducing the column temperature to below ambient. Vacuum conditions play several roles: i) increasing $CO_2$ purity in step 3, ii) facilitating the evaporation of water for moisture-driven $CO_2$ release, including the possibility of evaporation of saline water resources, and iii) evacuating the desorbed $CO_2$. After passing through a vapor vacuum blower in step 4, water vapor is cooled below its saturation temperature using the water reservoir as a heat sink, recovering the heat lost in the water reservoir during evaporation. This produces a liquid water stream that can be recycled as shown in Fig. 1 or captured as an upgraded product. Exact water tolerances and pressures of the final $CO_2$ gas stream are not considered here, but it is understood that different levels of water removal and compression pressure are required for supercritical $CO_2$ transported via pipeline versus a fuel/chemical $CO_2$ utilization pathway. In this study we assume $CO_2$ compression energetics up to 0.1 MPa (1 bar) and assume water leaving steps 4 and 5 are fully recovered. Considering its novelty and the central role of water management, VMS will require comprehensive modeling and optimization to evaluate its feasibility and performance for commercial DAC deployment.

In this work, we investigate the VMS process using a moisture-responsive sorbent, Amberlite IRA 900, ion-exchanged into a $CO_2$ reactive form. Building on recently developed bicomponent $CO_2 - H_2O$ isotherm models and experimentally measured enthalpies as well as kinetics for this material, we evaluate VMS performance on three key metrics: (1) $CO_2$ productivity, defined as the daily capture rate per unit sorbent mass ($\mathrm{kg\ kg_s^{-1}\ day^{-1}}$); (2) electrical energy demand per mass of $CO_2$ captured $\left(\mathrm{MJ \cdot kgCO_2}^{-1}\right)$; and (3) water loss arising from evaporation into the air stream during $CO_2$ sorption $\left(\frac{H_2O}{CO_2}\ \text{mass ratio}\right)$. We then apply an equal weight multi-objective optimization framework to quantify the tradeoffs among operating conditions that maximize $CO_2$ productivity while minimizing both energetic and water burdens across variable ambient humidity. We further examine the sensitivity of system performance to sorption kinetics. Finally, we compare the optimized VMS performance to leading DAC systems based on supported amine sorbents and hydroxide solvents.

# 2. Methods

## 2.1 Sorbent characterization

Amberlite IRA900 is a macroporous anion exchange resin composed of a styrene-divinylbenzene crosslinked matrix functionalized with trimethylammonium groups charge-balanced with chloride anions. This resin is characterized by tan, opaque spherical beads with particle diameters ranging between 640 - 800 $\mu m$. The resin's ion exchange capacity (IEC) is variable and commonly between 1.8-3 mol/kg. This embedded charge results in the resin's high water sorption capacity that can range from 58 to 64 wt%.[42]

### *2.1.1 Ion exchange*

For IRA900 to react with $CO_2$, it must first be ion exchanged from the chloride counter-ion to either the hydroxide, carbonate, or fully $CO_2$ loaded bicarbonate state. In the experimental studies of this work, IRA900 was ion exchanged into the bicarbonate form by submerging the resin into a 0.1M $KHCO_3$ solution for 24 hours, followed by a deionized water rinse to remove residual salt. This process was repeated to maximize the bicarbonate ion exchange.

*2.1.2 $CO_2$ and $H_2O$ isotherms*

In prior work, IRA900's bicomponent $CO_2/H_2O$ isotherm behavior was characterized over three different temperatures (15-35 ºC) and over pressures relevant to dilute $CO_2$ separation.[43] In that study, the bicarbonate ion exchange capacity was 1.91 mol/kg after three ion-exchange rinses. Pure water sorption exhibited a type II isotherm, indicative of multilayer adsoprtion on a porous material. This isotherm behavior was well described using the Guggenheim-Anderson-de Boer (GAB) model eqn (2).

$$q_{\mathrm{H_2O}} = \frac{q_m\, K_{ads} CRH}{(1-K_{ads}RH)(1+(c-1)K_{ads}RH)} \quad (2)$$

$$K_{ads} = \exp\left(\frac{E_{ml}-E_C}{RT}\right) \quad (3)$$

$$C = \exp\left(\frac{E_{mo}-E_C}{RT}\right) \quad (4)$$

Where $q_{\mathrm{H_2O}}$ is the loading of water, $q_m$ is an approximated monolayer loading, $RH$ represents the relative humidity or water activity, $K_{ads}[-]$ and $C[-]$ are temperature dependent affinity parameters, and $RT$ represents the product of the ideal gas constant and temperature.[44,45] In these constants (eqn (3-4)), the heats of sorption are approximated for a monoloayer of adsorbed water, $E_{mo}$ , and additional multilayers, $E_{ml}$, that are differenced relative to the heats of condensation for water$E_C$. The empirical temperature dependence of the enthalpys, as well as $H_2O$ and $CO_2$ isotherm parameters are given in .The dependence of $CO_2$ on $H_2O$ water sorption was neglected because $H_2O$ is an order of magnitude more soluble than $CO_2$ in this resin, and only minor deviations in water uptake were observed at low $CO_2$ oncentrations.[39]

However, it is water's influence on $CO_2$ sorption, $q_{\mathrm{CO_2}}$, that is responsible for the MS separation. This dependence was described using a modified Langmuir isotherm, where the impact of hydration on the $CO_2$ sorption mechanism was incorporated through a negative power law dependence on the water loading, $[q_{\mathrm{H_2O}}]^{-n}$, which in effect modifies the simple sorption equilibrium constant, $K_{MS}$ (eqn (5)).

$$q_{CO_2} = IEC\,\frac{K_{MS}\,[q_{\mathrm{H_2O}}]^{-n}\,P\mathrm{CO_2}}{1+K_{MS}\,[q_{\mathrm{H_2O}}]^{-n}P\mathrm{CO_2}} \quad (5)$$

In the above expression, the maximum loading was taken to be the IEC, as this is the greatest amount of chemisorbed $CO_2$ (as $HCO_3^-$) that can be loaded into the sorbent. Physisorption capacity was not considered given the low $CO_2$ partial pressures experienced in ambient air and in a vacuum-based MS process being explored in this study. The magnitude of the power exponent, *n*, is related to the stoichiometric water released upon drying and binding $CO_2$. Fig. 2(a) shows the $CO_2$ isotherm dependence on partial pressure at three different RH(x·100) (20, 50 and 80) evaluated at two temperatures 25℃ and 41℃. Fig. 2(b) Presents the $H_2O$ isotherm dependence on temperature (25℃ & 41℃ ) as a function of relative humidity.

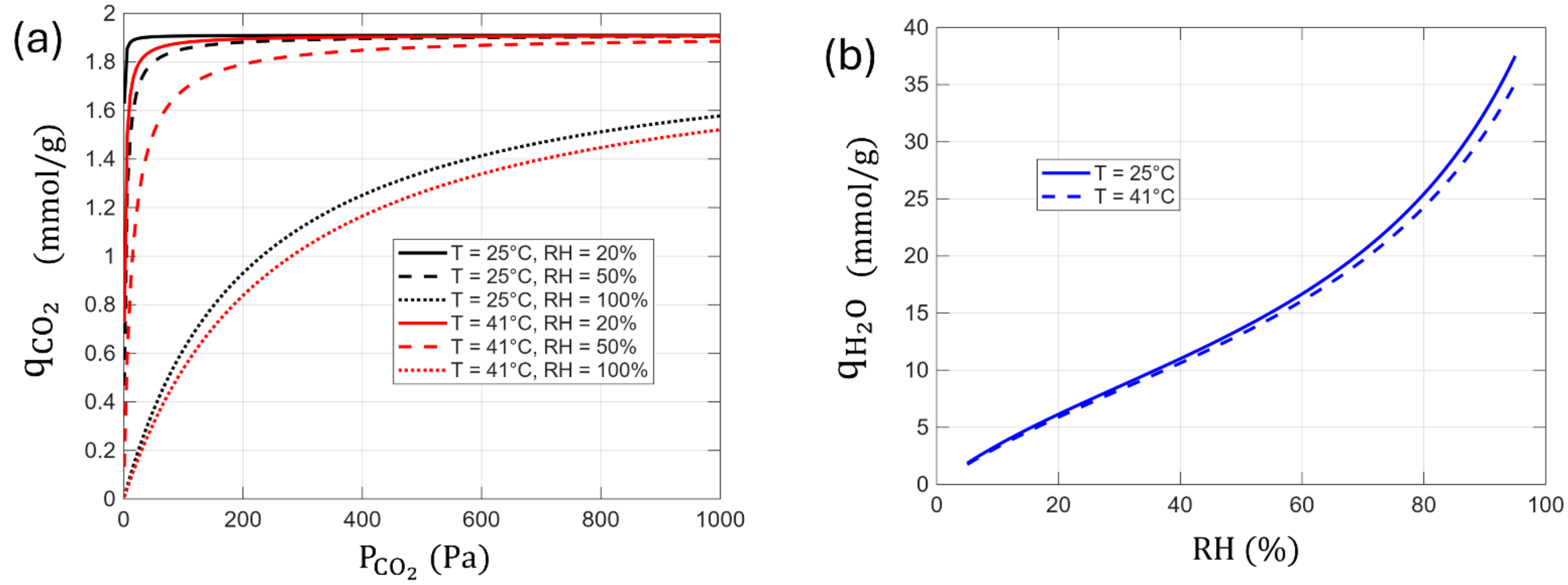

Fig. 2 species isotherm behavior where (a) $CO_2$ loading as a function of partial pressure for 20%, 50% & 80% RH at temperatures of 25℃ & 41℃ (b) shows $H_2O$ loading as a function of relative humidity at 25℃ and 41℃.

**Table 1 Isotherm parameters for water-dependent $CO_2$ and pure $H_2O$ isotherms**

| Parameter | Value | Units |
|---|---|---|
| **Modified Langmuir Model** | | |
| Ion exchange capacity, IEC | 1.91 | [mol kg$^{-1}$] |
| Equilibrium constant, $K_{MS}$ | 1x $10^{37}$exp $(-0.25 * T)$ | [kg mol$^{-1}$Pa$^{-1}$] |
| Water exponential factor, $n$ | 1400 exp $(-0.02 * T)$ | — |
| $CO_2$ Heats of sorption, $-\Delta H_{CO_2}$ | 70 | [kJ mol$^{-1}$] |
| **GAB model** | | |
| Monolayer loading capacity, $q_m$ | 11.4 | [mol kg$^{-1}$] |
| $H_2O$ Heats of sorption, $-\Delta H_{H_2O}$ | 46 | [kJ mol$^{-1}$] |
| **Temperature Dependent GAB Parameters** | | |
| Monolayer heats of sorption, $E_{mo} = -E * T + F$ | | [J mol$^{-1}$] |
| Multilayer heats of sorption, $E_{ml} = -G * T + H$ | | [J mol$^{-1}$] |
| Heats of condensation, $E_C = -J * T + Q$ | | [J mol$^{-1}$] |
| Fitted parameter in monolayer, $E$ | 38.6 | [J mol$^{-1}$ K$^{-1}$] |
| Fitted parameter in monolayer, $F$ | 59338 | [J mol$^{-1}$] |
| Fitted parameter in multilayer, $G$ | 50.4 | [J mol$^{-1}$ K$^{-1}$] |
| Fitted parameter in multilayer, $H$ | 58322 | [J mol$^{-1}$] |
| $J$ | 44.38 | [J mol$^{-1}$ K$^{-1}$] |
| $Q$ | 57220 | [J mol$^{-1}$] |

*2.1.3 Moisture swing kinetics*

Rates of combined $CO_2$ and $H_2O$ loading on ion-exchanged IRA900 were determined from gravimetric and integrated gas-phase composition data collected under controlled open-flow conditions. Experiments were performed *isothermally* using a Netzsch STA 449 F3 Jupiter thermogravimetric analyzer (TGA) coupled to a ProUmid MHG32 modular humidity generator and a LiCOR LI7000 infrared gas analyzer. The humidity generator was used to precisely adjust relative humidity to simulate a moisture-swing environment, as 200 sccm $CO_2/N_2$ mixtures (400 ppm $CO_2$ in ultra-high-purity $N_2$) were humidified to prescribed values and mixed before entering the TGA sample chamber. The sample temperature was maintained at 25 ± 0.3 °C at the location of the sample through a combination of a copper furnace and expansion gas cooling applied externally to the sample chamber walls.

The bead-form IRA900 samples (~8 mg) were spread in a single layer across shallow aluminum pans to ensure uniform exposure to the gas phase. Prior to the measurement of a step change in humidity, the sorbent was equilibrated in the sample chamber at 20% RH in 400 ppm $CO_2$ for ten hours to establish a reproducible $CO_2$ loaded state before testing. Sorption cycles consisted of sequential 5 h humidity steps (20↔95% RH), repeated three times, followed by a final 3 h dry $N_2$ purge to establish the dry sorbent reference mass.

Gravimetric changes were deconvoluted into $CO_2$ and $H_2O$ contributions ($q_{CO_2}, q_{H_2O}$) by combining the mass signal with real-time $CO_2$ concentration data from the gas analyzer. $CO_2$ loading was quantified by integrating the area under the gas concentration curves and converting this to molar flow rates, while $H_2O$ uptake was obtained from the residual mass difference after accounting for $CO_2$ (de)sorption. This combined approach enabled direct resolution of counter-sorption processes and partitioning of total sorbate loading between $CO_2$ and $H_2O$. Further details of the experimental setup and data analysis have been described in an earlier study.[39]

The overall kinetics of $CO_2$ and $H_2O$ sorption following a step change in water vapor concentration (20↔95% RH) were described using a linear drive force (LDF) expression as shown in eqn (6). In this formulation, $q_i$ represents the sorbed concentration of species $i$ at time, t, $q_i^*$ is the steady state loading, and $q_i(0)$ is the initial loading.

$$\frac{\partial q_i}{\partial t} = k_i(q_i^* - q_i) \tag{6}$$

Linearization of eqn (6) yields eqn (7), which allows determination of the mass transfer coefficient, $k_i$, by linear regression:

$$ln\frac{q_i(t) - q_i^*}{q_i(0) - q_i^*} = -k_i t \tag{7}$$

In the numerical process model, the assigned $k_i$ values depend on the prevailing sorption direction. For example, during $CO_2$ sorption concurrent with $H_2O$ desorption, $k_{CO_2}^{ads}$ and $k_{H_2O}^{des}$ were used to capture counter-sorption behavior. In the vapor-stripping step, where water vapor drives $CO_2$ release, the model employed $k_{CO_2}^{des}$ and $k_{H_2O}^{ads}$. In the final desorption, both species desorb simultaneously; their respective desorption values were applied.

## 2.2 Numerical model

### *2.2.1 Spatial domain*

In this work, we consider a packed bed sorption column that embodies a large cross-sectional diameter (D = 0.08m) relative to its length (L = 0.01m), as illustrated in Fig. 3 below (which is not drawn to scale but aims to represent the relation between the diameter and length). The low aspect ratio of column length and cross-section is chosen by most air contactor systems to minimize the pressure drop energetics when pushing large volumes of air to extract dilute concentrations of $CO_2$. This configuration was selected to align with the experimental setup reported by Wurzbacher *et al.*[44] and to ensure consistency with previous amine-based TVSA studies used for comparison.[23,29] The governing transport equations in the packed bed are derived based on the following assumptions:

- The gas phase flow is modeled as an axially dispersed plug flow, where flow in the radial direction is ignored.
- The walls of the sorbent column are approximated as adiabatic.
- The gas phase follows the ideal gas law, and ambient air is approximated as nitrogen, $N_2$, where nitrogen sorption into the sorbents is assumed negligible.
- The sorption kinetics are described using the LDF model.
- The pressure drop along the bed is approximated by the Ergun equation.
- The sorbent properties and bed voidage are considered constant and uniform throughout the column.
- The water source for vapor stripping is assumed sufficiently large to remain isothermal, thus evaporative cooling in the water reservoir is neglected.
- A modified water vapor-liquid equilibrium equation (e.g. Antoine's equation) is used to represent water activity equivalent to sea water.

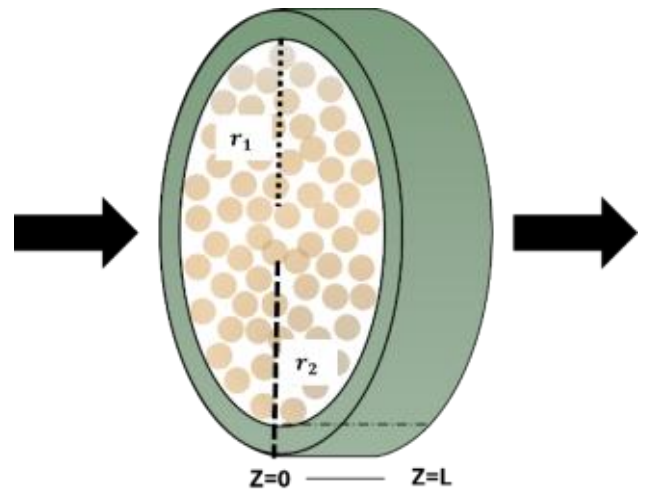

**Fig. 3** Schematic of a packed-bed sorption column showing gas species flow from the inlet to the outlet, with inner and outer radii highlighted

The transport model for gas species in both the bulk gas phase and their source/sink due to (de)sorption into the solid phase are defined by the conservation of mass, chemical species and energy with the governing conservation equations detailed below. All variables used in each equation are defined in the nomenclature in the appendix.

*2.2.2 Governing equations*

**Mass Balance:**

The total mass balance in the sorption column is given as

$$\frac{\partial P}{\partial t} = \left(-T\frac{\partial}{\partial z}\left(\frac{Pv}{T}\right)\right) + \left(\frac{P}{T}\frac{\partial T}{\partial t}\right) - \left(\frac{(1-\varepsilon_b)}{(\varepsilon_b)}RT\sum_{i=1}^{n_{comp}}\frac{\partial q_i}{\partial t}\right) \quad (8)$$

Where $P$ is the total pressure, $z$ is the spatial coordinate along the column, $v$ is the interstitial velocity, $\varepsilon_b$ is the bed voidage, $R$ is the universal gas constant, $T$ is the temperature, and $q_i$ is the adsorbed phase concentration of component $i$. This formulation couples the gas-phase and solid-phase behavior. The first term captures gas-phase mass transfer due to convective flux along the column, the second term links pressure dynamics to transient temperature variations, and the third term captures the source/sink of gases due to (de)sorption.

**Pressure Drop:**

The pressure drop is approximated by the empirical Ergun equation, which accounts for viscous and inertial dissipation of mechanical energy in packed beds.

$$-\frac{\partial P}{\partial z} = \beta v_{int} + \alpha v_{int}^2 \quad (9)$$

The coefficients $\beta = \frac{150}{4}\frac{1}{r_p^2}\frac{(1-\varepsilon_b)^2}{\varepsilon_b^2}\mu$ and $\alpha = 1.75\frac{(1-\varepsilon_b)}{\varepsilon_b}\frac{\rho_g}{r_p}$ represent the viscous and kinetic effects, where $r_p$, $\mu$ and $v_{int}$ are the particle radius, viscosity and interstitial velocity, respectively. $\rho_g$ is the gas density defined by the ideal gas law $\rho_g = \frac{P\sum MW_i}{RT}$, where $MW_i$ is the molecular weight of a species $i$.

The Ergun equation is commonly expressed with the superficial velocity, which refers to the velocity of fluid flowing through the entire cross-sectional area of the bed.[46] Modeling a packed bed with sorbent particles requires a velocity that accounts for the void spaces. Hence, the superficial velocity in Ergun's equations is re-written as interstitial velocity. The relationship between the two is given as: $\mathrm{v_{sup}} = \mathrm{v_{int}} * \varepsilon_\mathrm{b}$.

**Chemical Species Balance:**

The conservation of species considers the transformation of $CO_2$ and $H_2O$ as these molecules flow through the sorbent bed and (de)sorb between the gas phase and solid sorbent. The third component, $N_2$, is handled as the difference between the total gas pressure and the partial pressures of $CO_2$ and $H_2O$. The chemical species balance is given in eqn (10)

$$\frac{\partial c_i}{\partial t} = D_L \frac{\partial^2 c_i}{\partial z^2} - \frac{\partial}{\partial z}(c_i v) - \frac{1-\varepsilon_b}{\varepsilon_b} * \frac{\partial q_i}{\partial t} \tag{10}$$

$\frac{\partial c_i}{\partial t}$ represents the accumulation or depletion of species over time. $D_L \frac{\partial^2 c_i}{\partial z^2}$ is the diffusion term with an axial mixing diffusion coefficient ($D_L$), $\frac{\partial}{\partial z}(cv)$ is the advection term that defines the transport based on the fluid's axial velocity and $\frac{\partial q_i}{\partial t}$ is the source or sink of rate of sorption.[47] A simplified equation describing axial dispersion in a packed bed, based on the effects of molecular diffusion and turbulent mixing, is given below in eqn (11)

$$D_L = 0.7 D_m + 0.5\, v d_p \tag{11}$$

$D_m$ is the molecular diffusivity and $d_p$ is the particle diameter.

**Energy Balance:**

In a one-dimensional reactor column, the transient energy balance, eqn (12), accounts for the energy stored in the sorbent ($\rho_s C_{p,s}$) and energy stored in the adsorbed phase ($\sum_{i=1}^{n_{comp}} C_{p,a_i} q_i$). Axial heat transfer occurs through conduction along the bed $\left(\frac{K_z}{\varepsilon_b}\frac{\partial^2 T}{\partial z^2}\right)$, and advection by the flowing gas $\left(\frac{C_{p,g}}{R} * \frac{\partial}{\partial z}(vP)\right)$. Transient gas compression and expansion effects are captured by $\left(\frac{C_{p,g}}{R} * \frac{\partial P}{\partial t}\right)$. The temperature-dependent changes in adsorbed loading are denoted by the term $\left(\frac{1-\varepsilon}{\varepsilon} T \sum_{i=1}^{n_{comp}} C_{p,a_i} \frac{\partial q_i}{\partial t}\right)$, and the source/sink terms due to enthalpy of (de)sorption are represented by $\frac{1-\varepsilon}{\varepsilon} \sum_{i=1}^{n_{comp}} \left((-\Delta H_i)\frac{\partial q_i}{\partial t}\right)$.

$$\left[\frac{1-\varepsilon_b}{\varepsilon_b}\left(\rho_s C_{p,s} + \sum_{i=1}^{n_{comp}} C_{p,a_i} q_i\right)\right]\frac{\partial T}{\partial t} = \left(\frac{K_z}{\varepsilon_b}\frac{\partial^2 T}{\partial z^2}\right) - \left(\frac{C_{p,g}}{R} * \frac{\partial}{\partial z}(vP)\right) - \left(\frac{C_{p,g}}{R} * \frac{\partial P}{\partial t}\right) - \left(\frac{1-\varepsilon_b}{\varepsilon_b} T \sum_{i=1}^{n_{comp}} C_{p,a_i} \frac{\partial q_i}{\partial t}\right) + \frac{1-\varepsilon_b}{\varepsilon_b} \sum_{i=1}^{n_{comp}} \left((-\Delta H_i)\frac{\partial q_i}{\partial t}\right) \tag{12}$$

In the above expression, $C_{p,s}$ is the specific heat capacity of the sorbent, $C_{p,g}$ is the specific heat capacity of the gas phase, $C_{p,a_i}$ is the specific heat capacity of the sorbed phase of each specie $i$, $\rho_s$ is the sorbent density, $\Delta H_i$ is the enthalpy of (de)sorption of the species $I$, and $K_z$ is the thermal conductivity. All these constants can be found in **Table 2**.

*2.2.3 Numerical approach*

The governing partial differential equations (PDEs) listed above are solved numerically in their dimensional form using the finite volume method (FVM). The column is treated as one-dimensional (1D) problem and discretized into N control volumes with uniform spacing, $\Delta z$. Each control volume center is indexed by $j$, with boundaries located at $j \pm 0.5$ as shown in Fig. 4. The column inlet and outlet boundaries are represented by 0.5 and $N + 0.5$ respectively. In this 1D finite volume mesh, fluxes in the transport equations are evaluated at the cell faces ($j \pm 0.5$), while all other properties are defined at the cell centers, $j_i$.

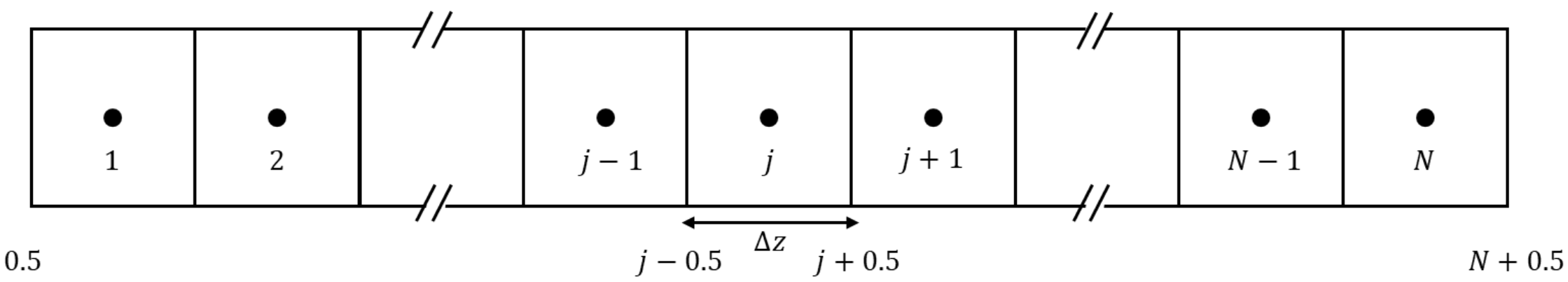

**Fig. 4** Schematic of the sorption column and its finite volume discretization used for numerical modeling.

Calculating fluxes at the cell walls requires state variables at $j-1$, $j$ and $j+1$, which are available for internal cells ($j = 2, \ldots., N-1$). Rearranging eqn (9) yields the discretized expression for the velocities at the cell walls:

$$v_{j+0.5} = \frac{\beta \pm \sqrt{\beta^2 - 4\alpha\left(\frac{P_{j+1} - P_j}{\Delta z}\right)}}{2\alpha} \tag{13}$$

Evaluating flux at the left and right boundaries requires values outside the domain. To avoid introducing a ghost cell, a first order accurate half-cell approximation, $f_1 - f_0 = 2(f_1 - f_{0.5})$, is applied The resulting left ($z = 0.5$) and right ($z = N + 0.5$) velocity boundary conditions are as follows:

$$v_{0.5} = v|_{z=0} = \frac{\beta \pm \sqrt{\beta^2 - 4\alpha\left(\frac{P_1 - P_{0.5}}{\Delta z/2}\right)}}{2\alpha} \tag{14}$$

$$v_{N+0.5} = v|_{z=L} = \frac{\beta \pm \sqrt{\beta^2 - 4\alpha\left(\frac{P_{N+0.5} - P_N}{\Delta z/2}\right)}}{2\alpha} \tag{15}$$

For cases where the inlet or outlet velocities are prescribed ($v_{0.5}$ $and$ $v_{N+0.5}$), such as during $CO_2$ sorption and vapor stripping, eqn (14) and eqn (15) can be re-arranged to find the inlet (eqn (16)) and outlet pressures (eqn (17)).

$$P_{0.5} = P|_{z=0} = P_1 + \frac{\Delta z}{2}(\beta v_{0.5} + \alpha v_{0.5}^2) \tag{16}$$

$$P_{N+0.5} = P|_{z=L} = P_N - \frac{\Delta z}{2}(\beta v_{N+0.5} + \alpha v_{N+0.5}^2) \tag{17}$$

The pressure boundaries in pressurization (inlet), air evacuation (outlet), and final desorption (outlet) are defined using a first-order exponential relaxation expression with a pressure relaxation factor of $\lambda = 0.5$ and are presented as eqn (18), eqn (19) and eqn (20) respectively.

$$P_{0.5} = P|_{z=0} = P_{atm} - (P_{atm} - P_L)e^{-\lambda t_{PZ}} \tag{18}$$

$$P_{N+0.5} = P|_{z=L} = P_I + (P_{N+0.5} - P_I)e^{-\lambda t_{EV}} \tag{19}$$

$$P_{N+0.5} = P|_{z=L} = P_L + (P_{N+0.5} - P_L)e^{-\lambda t_{DES}} \tag{20}$$

Where $P_{atm}$ is the atmospheric pressure, $P_L$ is the target pressure for step 5, $P_I$ is the intermediate pressure during step 3 and $t_{PZ}$, $t_{EV}$, $t_{DES}$ represent the duration of each step (1, 3 and 5). This first-order exponential relaxation formulation captures the transient pressure dynamics by simulating a gradual increase or decay from the current column pressure to the respective target pressure. This boundary treatment avoids abrupt pressure transitions and improves numerical stability. Because these are significantly shorter cycles compared to $CO_2$ sorption (step 2) and vapor stripping (step 4), the use of this factor has negligible impacts on energy calculations as described in section 2.2.5. All other boundary conditions for each of the dependent parameters used by the VMS model are also provided in the supplementary information, Table S1.

The Weighted Essentially Non-Oscillatory (WENO) finite volume scheme of the governing equation advection terms (velocity on the walls) was used to minimize numerical dispersion of sharp pressure and concentration gradients, following the approach proposed by Haghpanah.[48] The resulting 1D discretized system is integrated in time using MATLAB's ODE15s solver to efficiently handle the stiffness of the governing PDEs. Further details on the implementation of this numerical scheme may be found in the referenced work.[48–50]

### *2.2.4 Modeling VMS cycles*

A simulated VMS cycle begins after the completion of a $CO_2$ desorption step, entering a pressurization stage where the column is brought to ambient pressures using air (Fig. 1). In this pressurization step, feed gas enters the inlet at ambient pressure, $P_{atm} = 100kPa, T = 298\ K$, 400 ppm $CO_2$ and a prescribed RH while the outlet remains closed, rapidly increasing the column pressure for a duration of $t_{PZ}$. Once pressurized, the system transitions into the $CO_2$ sorption step, in which both the inlet and outlet are open. The same feed air passes through the column for a duration, $t_{CS}$, at a fixed and prescribed velocity, $v_{AIR}$, which in turn dictates the pressure drop across the bed, eqn (9). After the $CO_2$ sorption step, the packed bed moves into the air evacuation stage, where the inlet is closed, and a vacuum is applied to the outlet drawing down the pressure to an intermediate value of $P_I = 0.5\ kPa$. Purging the column to $P_I$ removes nitrogen and oxygen, increasing $CO_2$ purity in the downstream recovery steps. The evacuation pressure must be low enough to remove excess air while avoiding early desorption of $CO_2$.

Following air evacuation is vapor stripping (step 4, Fig. 1), which plays a central role in desorbing the $CO_2$ in the VMS process. With both the inlet and outlet open, water vapor is pulled through the column via the vacuum placed at the outlet. The inlet pressure corresponds to the saturation pressure of water at $T = 300K$, while the outlet is prescribed according to a fixed velocity, $v_{VS}$, again dictating the pressure drop. The vacuum pressure facilitates the vaporization of water, which binds to the sorbent upon passing through the sorbent bed. Sorbed water leads to $CO_2$ desorption in the MS sorbents as described by the reverse of eqn (1). Further, the flow of water vapor sweeps $CO_2$ out of the sorbent bed, dropping the bed $PCO_2$ which leads to a deeper unloading of $CO_2$ and increased productivity of the separation.

Finally, a desorption step is performed by closing the water vapor inlet and applying a vacuum at the outlet pressure with a target pressure, $P_L = 1\ kPa$. This step removes residual $CO_2$ and recovers water vapor from the gas phase, accompanied by a temperature decrease due to evaporative cooling. The resulting temperature drop restores the sorbents affinity for $CO_2$, resetting the column for the next $CO_2$ sorption phase.

The duration of each step that invokes a step change in pressure (pressurization and air evacuation) was run until target pressures were attained. In step 2 ($CO_2$ sorption), step 4 (vapor stripping) and step 5 (final desorption), the duration was optimized along with the air or vapor flow rate, respectively (Section 3.3).

In the baseline simulation, the sorbent bed was initialized uniformly with 400 ppm $CO_2$, 20-80% RH, $T = 298\ K$ and a total pressure $P_L = 1\ kPa$, balanced with $N_2$. The steps were then cycled until cyclic steady state (CSS) conditions were reached, which are defined using two convergence criteria. The first CSS criterion requires that $CO_2$ mass balance over a full cycle be within 0.5% as indicated in eqn (21) below:

$$CSS_1 = \frac{|CO_{2,out} - CO_{2,in}|}{CO_{2,in}} x\ 100 < 0.5\% \tag{21}$$

Where $CO_{2,\ out}$ represents the total $CO_2$ exiting the system during steps 2, 3, and 4, while $CO_{2,\ in}$ represents the total $CO_2$ entering the system during steps 1 and 2 in a single cycle. The second criterion is satisfied when the normalized difference of each state variable across all internal nodes (i.e. pressure, temperature, species mole fraction and species solid phase loading) between successive cycles is within 0.5% as defined in eqn (22).

$$\Delta s_{i,j} = \left| \frac{s_{i,j}^{final} - s_{i,j}^{intial}}{s_{i,j}^{initial}} \right| \quad for\ j = 2,3,\ldots,N+1 \tag{22}$$

Here $s_{i,j}^{final}$ is the value of state variable $i$ at node $j$ in the current cycle and $s_{i,j}^{initial}$ is the value of state variable $i$ at node $j$ in the previous cycle. A simulation was deemed to have reached CSS when both criteria had been met for five consecutive cycles. All of the sorbent bed and gas phase thermodynamic and transport properties used by the VMS model are listed below in **Error! Reference source not found.**.

**Table 2 Parameters and properties used in the baseline simulation model**

| Parameter | Value |
|---|---|
| **Bed Properties** | |
| length, $L$ [m] | 0.01 |
| inner column radius, $r_{in}$ [m] | 0.040 |
| outer column radius, $r_{out}$ [m] | 0.041 |
| bed voidage, $\varepsilon_b$ [-] | 0.35 |
| particle void $\varepsilon_p$ [-] | 0.24 |
| particle radius, $r_p$ [m] | 4x$10^{-4}$ |
| tortuosity, $\tau$ [-] | 3 |
| | |
| **Constant properties** | |
| adsorbent density $\rho_s$ [kg $m^{-3}$] | 1060 |
| bed wall density $\rho_w$ [kg $m^{-3}$] | 7800 |
| specific heat capacity of gas phase, $C_{p,g}$ [J $mol^{-1} k^{-1}$] | 29 |
| specific heat capacity sorbed $CO_2$, $C_{p,a,CO_2}$ [J $mol^{-1} k^{-1}$] | 37 |
| specific heat capacity of sorbed water, $C_{p,a,H_2O}$ [J $mol^{-1} k^{-1}$] | 75 |
| specific heat capacity of adsorbent, $C_{p,s}$ [J $kg^{-1} k^{-1}$] | 1580 |
| fluid viscosity, $\mu$ [kg $m^{-1} s^{-1}$] | 1.72 x$10^{-5}$ |
| molecular diffusivity, $D_m$ [$m^2$ $s^{-1}$] | 2.05 x$10^{-5}$ |
| gas thermal conductivity $K_z$ [J $m^{-1} k^{-1} s^{-1}$] | 0.09 |
| thermal conductivity, $K_w$ [J $m^{-1} k^{-1} s^{-1}$] | 16 |
| gas constant, $R$ [ Pa $m^3$ $mol^{-1} k^{-1}$] | 8.314 |
| molecular weight of air, $MW_{AIR}$ [kg $mol^{-1}$] | 0.02896 |
| molecular weight of water, $MW_{H_2O}$ [kg $mol^{-1}$] | 0.01805 |
| adiabatic constant, $\gamma$ [$-$] | 1.4 |
| turbomachinery efficiency, $\eta$ [$-$] | 0.72 |
| pressure rate constant, $\lambda$ [$-$] | 0.5 |

To ensure the reliability of the VMS numerical model, we based our initial code on an open-source two stage PVSA simulation[51] by the team of Karson *et al*[52] and validated it against the experimentally verified data of Haghpanah *et al.*[48], who simulated a PVSA process capturing $CO_2$ from a flue gas powerplant. By running simulations under matching PVSA conditions, we successfully reproduced the reported results, confirming that our base model accurately captures the transient mass and heat transfer behaviors. Mass and energy balances were also checked to ensure consistency with the reported measurements. Following this verification process, the base model was adapted for the unique VMS process, incorporating experimentally determined $CO_2/H_2O$ isotherms and kinetics.

*2.2.5 Performance Metrics*

Upon reaching CSS, the performance of the VMS cycle was evaluated using several key performance metrics. These included: (1) the productivity per cycle, based on the $CO_2$ captured in steps 4-5; (2) the mechanical energy expended during steps 2-5; (3) the $H_2O$ loss through evaporation in air during step 2; (4) the quantity of $H_2O$ processed in step 4, with the majority recycled from water knockout during steps 4-5; and (5) the fraction of $CO_2$ captured in step 4-5 to the $CO_2$ adsorbed in step 1-2; and. Together, these metrics form the basis for assessing and optimizing the cycle's operational efficiency.

Productivity is defined as the mass of $CO_2$ captured from air per unit mass of sorbent per day and computed in eqn (23).

$$\mathrm{CO_2\ Productivity}\left[\frac{\mathrm{kgCO_2}}{\mathrm{kg_S \cdot day}}\right] = \frac{\text{mass of CO}_2\text{in the extract stream per cycle}}{\text{(mass of sorbent)(day)}} \tag{23}$$

This productivity metric does not consider any $CO_2$ generation associated with the electrical energy required for operation. Those contributions are more appropriately addressed within full technoeconomic and life-cycle analysis, which depends on the carbon intensity of the electrical infrastructure (e.g. regional grid, solar + storage, etc.). This definition of $CO_2$ productivity is used consistently throughout this study and is compared directly with reported values from leading DAC technology studies in Section 3.5.

The electrical energy requirement in the VMS system arises from gas and vapor transport through the column during $CO_2$ sorption (eqn 24), vacuum pumping of residual air (eqn 25) and water vapor (eqns 26,27) in steps 3-5, and compressing the captured $CO_2$, along with any air impurities, to atmospheric pressure (eqn 28). Turbomachinery power consumption is calculated assuming adiabatic gas compression with a prescribed thermodynamic efficiency, $\eta$, set to 72% for all machine types.

The following equations were used to compute energy consumption (En) for each step:

**$CO_2$ Sorption**

$$\mathrm{En_{CS}} = \frac{1}{\eta}\frac{\gamma}{\gamma-1}\int_0^{t_{CS}} \sum \dot{n}_{i_{in}}\, RT_{in} \left[\left(\frac{P_{in}}{P_{atm}}\right)^{\frac{\gamma-1}{\gamma}} - 1\right] dt \quad (24)$$

**Air Evacuation**

$$\mathrm{En_{EV}} = \frac{1}{\eta}\frac{\gamma}{\gamma-1}\int_0^{t_{EV}} \sum \dot{n}_{i_{out}}\, RT_{out} \left[\left(\frac{P_{atm}}{P_{out}}\right)^{\frac{\gamma-1}{\gamma}} - 1\right] dt \quad (25)$$

**Vapor Stripping ($H_2O$)**

$$\mathrm{En_{VS,H_2O}} = \frac{1}{\eta}\frac{\gamma}{\gamma-1}\int_0^{t_{VS}} \dot{n}_{H_2O} RT_{out} \left[\left(\frac{P_{sat}(T_{sat})}{P_{out}}\right)^{\frac{\gamma-1}{\gamma}} - 1\right] dt \quad (266)$$

**Final Desorption ($H_2O$)**

$$\mathrm{En_{DES}} = \frac{1}{\eta}\frac{\gamma}{\gamma-1}\int_0^{t_{DES}} \sum \dot{n}_{i_{out}}\, RT_{out} \left[\left(\frac{P_{sat}(T_{sat})}{P_{out}}\right)^{\frac{\gamma-1}{\gamma}} - 1\right] dt \quad (277)$$

**$CO_2$ Compression to 1 atm**

$$\mathrm{En_{Comp}} = \frac{1}{\eta}\frac{\gamma}{\gamma-1}\int_0^{t_{VS}} \dot{n}_{CO_2+N_2} RT_{out} \left[\left(\frac{P_{atm}}{P_{out}}\right)^{\frac{\gamma-1}{\gamma}} - 1\right] dt\ +$$

$$\frac{1}{\eta}\frac{\gamma}{\gamma-1}\int_0^{t_{DES}} \dot{n}_{CO_2+N_2} RT_{out} \left[\left(\frac{P_{atm}}{P_{out}}\right)^{\frac{\gamma-1}{\gamma}} - 1\right] dt \quad (28)$$

In eqn (24-27), $\gamma$ denotes the adiabatic expansion coefficient, $\eta$, is the efficiency of the relevant turbomachinery (blower, vacuum pump, compressor), $\dot{n}_i$ is the molar flow rate of species $i$, $P_{sat}(T_{sat})$ is the saturation pressure of water at temperature $T_{sat}$. $P_{in}$, $T_{in}$, $P_{out}$, $T_{out}$ are the column inlet and outlet pressure and temperature respectively. No energy is expended during pressurization, as ambient air expands spontaneously into the column.

During vapor stripping, $CO_2$ and any residual air are compressed up to ambient pressure (eqn 28), while $H_2O$ vapor is compressed only to its saturation pressure (eqn 27), at which point it condenses to liquid water. This condensation occurs downstream of the vacuum pump (Fig. 1) and upstream of $CO_2$ compression. The heat of condensation is not explicitly calculated, as this energy is assumed to be exchanged with the water reservoir, effectively recovering the thermal energy lost during evaporation.

The time-dependent molar flow rates, $\dot{n}_i$, were determined based on velocity, mole fraction, pressure, and temperature at the corresponding inlet or outlet column position, using the ideal gas law:

$$\dot{n}_{i\,in} = \frac{\varepsilon_b A}{R} v_{0.5}(t) \frac{y_{i_{feed}} P_{0.5}(t)}{T_{0.5}(t)} \quad (28)$$

$$\dot{n}_{i_{out}} = \frac{\varepsilon_b A}{R} v_{N+0.5}(t) \frac{y_{i_{N+0.5}} P_{N+0.5}(t)}{T_{N+0.5}(t)} \quad (29)$$

$$\dot{n}_{CO_2+N_2} = \frac{\varepsilon_b A}{R} v_{N+0.5}(t) \frac{P_{N+0.5}(t)}{T_{N+0.5}(t)} \left( y_{CO_2\,N+0.5} + y_{N_2\,N+0.5} \right) \quad (30)$$

$$\dot{n}_{H_2O} = \frac{\varepsilon_b A}{R} v_{N+0.5}(t) \frac{y_{H_2O_{N+0.5}} P_{N+0.5}(t)}{T_{N+0.5}(t)} \quad (31)$$

The total energy consumption per mass of $CO_2$ captured is calculated in eqn (32) as

$$E_T \left[\frac{MJ}{kgCO_2}\right] = \frac{En_{CS} + En_{EV} + En_{VS} + En_{DES} + En_{Comp}}{\text{mass of } CO_2 \text{ in the extract stream per cycle}} \quad (32)$$

Since water management is a crucial aspect in the VMS process, we quantify water loss through evaporation in air during the $CO_2$ sorption step. The total water lost per mass of $CO_2$ captured is calculated (eqn (33)) as

$$\text{water loss} \left[\frac{kgH_2O}{kgCO_2}\right] = \frac{mole_{out,H_2O|CS} - (mole_{in,H_2O|PZ} + mole_{in,H_2O|CS})}{mole_{out,CO_2|VS} + mole_{out,CO_2|DES}} \frac{MW_{H_2O}}{MW_{CO_2}} \quad (33)$$

Further, the total water processed that includes water recovered from the downstream condenser of steps 4 and 5 (Fig 1) is calculated eqn (34) as

$$\text{water processed} \left[\frac{kgH_2O}{kgCO_2}\right] = \frac{mole_{in,H_2O|CS}}{mole_{out,CO_2|VS} + mole_{out,CO_2|DES}} \frac{MW_{H_2O}}{MW_{CO_2}} \quad (34)$$

Productivity, energy consumption and water loss are identified as principal optimization objectives for the VMS process, as they directly influence its technical feasibility, economic viability, and scalability for DAC applications. Achieving high productivity with minimal energy input and water loss is critical for reducing operational costs and environmental footprint while ensuring efficient $CO_2$ capture over shorter cycle times.

Although not a primary optimization target, $CO_2$ purity of the extract stream was tracked to evaluate the effectiveness of gas separation. Purity is calculated in eqn (35) as the sum of $CO_2$ mole fraction captured during vapor stripping and final desorption relative to the total moles of $CO_2$ and $N_2$ in the extract stream of the same steps. Water is not included in this calculation as we assume complete condensation of water downstream, as represented in Fig. 1.

$$\text{purity}(\%) = \left[\frac{mole_{out,CO_2|VS} + mole_{out,CO_2|DES}}{mole_{out,CO_2|VS} + mole_{out,CO_2|DES} + mole_{out,N_2|VS} + mole_{out,N_2|DES}}\right] \text{x } 100 \quad (35)$$

The $CO_2$ capture efficiency was also tracked to evaluate the overall effectiveness of the vapor stripping and final desorption processes and how these recover the loaded $CO_2$ during pressurization and sorption. This is calculated in eqn (36) as follows:

$$CO_2 \text{ captured } (\%) = \left[\frac{\text{mole}_{\text{out},CO_2|VS} + \text{mole}_{\text{out},CO_2|DES}}{\text{mole}_{\text{in},CO_2|PZ} + \text{mole}_{\text{in},CO_2|CS}}\right] \text{x } 100 \quad (36)$$

While water recovery was not an explicit optimization objective, it was tracked to quantify the amount that could be recycled to the feed or used as an upgraded water byproduct. The water upgrade occurs inherently through vacuum evaporation in the vapor stripping step. Water recovered is defined in eqn (37) as the moles of water recovered during vapor stripping and desorption in relation to the water processed in vapor stripping.

$$\text{water recovered } (\%) = \left(\frac{\text{mole}_{\text{out},H_2O|VS} + \text{mole}_{\text{out},H_2O|DES}}{\text{mole}_{\text{in},H_2O|VS}}\right) \text{x } 100 \quad (37)$$

## 2.3 Optimization Strategy

Optimization of the VMS separation was informed by insights from the baseline VMS case study reported in Section 3, with the primary objectives being to minimize energy consumption, minimize water loss, and maximize $CO_2$ productivity. The selected decision variables were $CO_2$ sorption time ($t_{CS}$), vapor stripping time ($t_{VS}$),, desorption time ($t_{DES}$), air velocity ($v_{AIR}$), and vapor stripping velocity ($v_{VS}$). These variables were chosen based on their significant impact on the energy-water loss-productivity trade-off. For example, longer sorption and vapor stripping times increase $CO_2$ capture but also raise blower and vacuum energy demands as well as increase water processed and lost. Optimizing velocities was especially critical, as it directly affects pressure drop across the system impacting the mechanical loads (eqn (24-27)). Although our case study indicated that a vapor stripping duration of twice the $CO_2$ sorption duration was beneficial for productivity, a balance was needed to stay within reasonable cycle durations.

**Table 3 Optimization decision variable workspace**

| | $t_{CS}$ [s] | $t_{VS}$ [s] | $t_{DES}$ [s] | $v_{AIR}$ [m/s] | $v_{VS}$ [m/s] |
|---|---|---|---|---|---|
| Lower bound | 1000 | 1000 | 200 | 0.1 | 0.01 |
| Upper bound | 10800 | 10800 | 500 | 1 | 5 |

Optimization was carried out using the non-dominated sorting genetic algorithm (NSGA-II) from MATLAB's global optimization toolbox. A population size of 10 times the decision variables and generation size of 30 were used to ensure diversity and adequate convergence. **Table 3** outlines the range used for each decision variable.

# 3. Results and discussion

## 3.1 Moisture swing kinetics

To evaluate sorbent performance in a sorption process, both thermodynamic and kinetic parameters must be considered. While prior work by Marques-Lopez *et al.* and Najaf-Tomaraei *et al*. has established bicomponent isotherm data in IRA900, here we employ open-flow counter-sorption experiments of $CO_2$ and $H_2O$ to extract first-order kinetic constants for each species.[43,53] The resulting cumulative uptake of $CO_2$ and $H_2O$ during moisture-swing counter sorption at 25 °C in 400 ppm $CO_2$ are presented in Fig. 5(a) and (c) correspond to $CO_2$ and $H_2O$ sorption behavior, respectively, from a decrease in humidity (95% → 20% RH). Under these dehydrating conditions, $CO_2$ binds to the sorbent, reaching a loading of 0.9 ± 0.1 mmol/g after around two hours while $H_2O$ desorption reaches a steady-state net loading of 24 ± 2 mmol/g after around one hour. Uncertainties represent the standard deviation from three MS cycles. Conversely, Fig. 5(e) and (g) correspond to $CO_2$ and $H_2O$ sorption behavior, respectively, from an increase in humidity (20% → 95% RH). It takes nearly three hours for both $CO_2$ desorption (1.1 ± 0.2 mmol/g) and $CO_2$ sorption (23 ± 1 mmol/g) to reach steady-state levels under the flow conditions of the measurement. We note that the loadings shown in Fig. 5 do not represent equilibrium isotherm loadings, as complete unloading of $CO_2$ and $H_2O$ was not achieved in these experiments. Instead, we quantified the change in loading associated with the 20→95% water vapor swing, commonly referred to as the working capacity. Furthermore, irrespective of sorption direction (i.e. loading or unloading), the change in sorption was evaluated as an absolute value relative to the initial condition of the water vapor step change. This approach enables clearer visualization of kinetic differences between sorption steps and accounts for variations in the total extent of (de)sorption observed across the three experimental cycles.

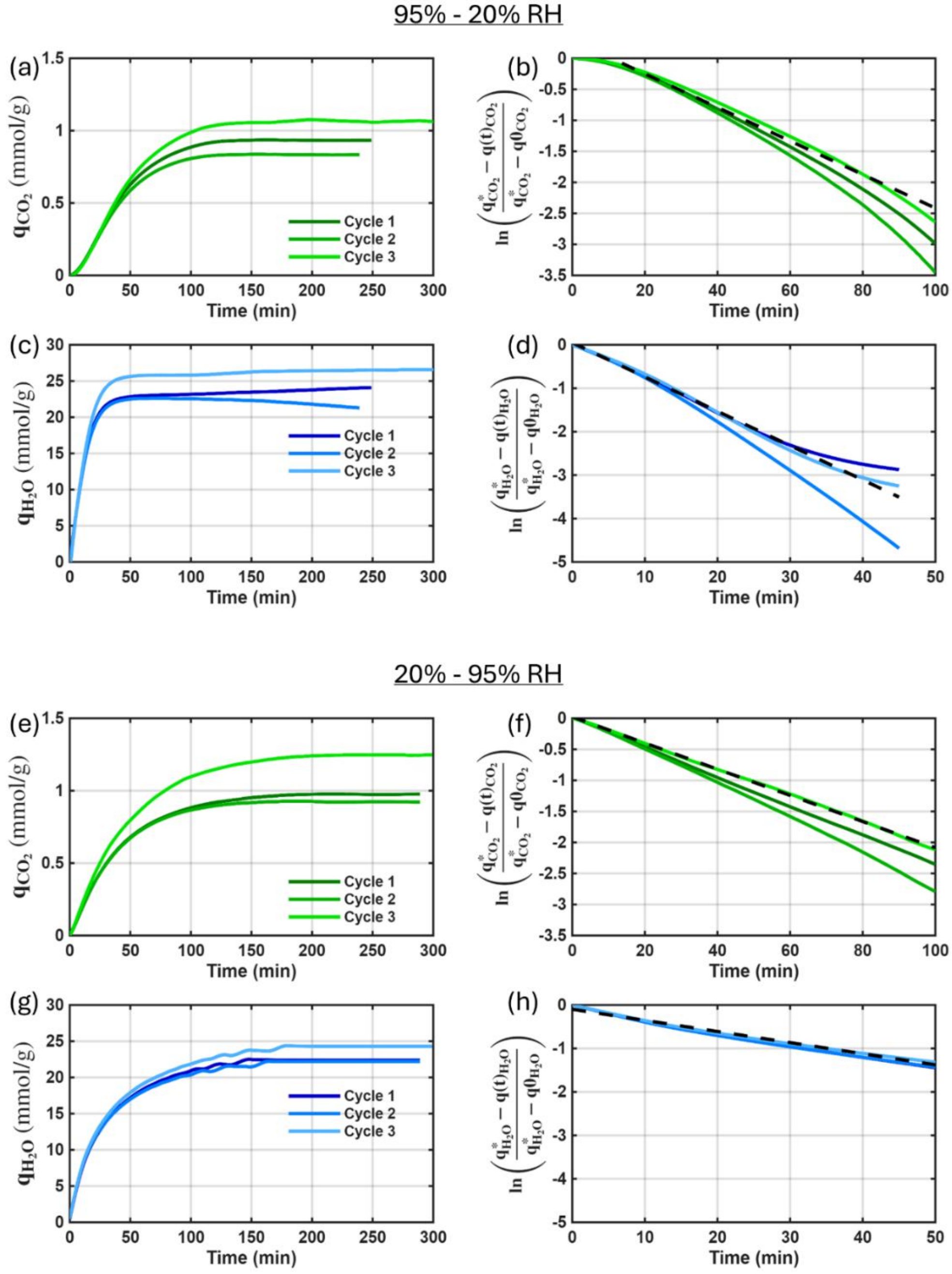

Fig. 5 Cumulative and linearized kinetic fits for $CO_2$ (green) and $H_2O$ (blue) loading on IRA900 measured via isothermal TGA (25ºC) with gas analysis over three cycles. For both 95% to 20% and 20% to 95% step changes in humidity, subplots presents results for (a), (e) $CO_2$ cumulative loading with corresponding (b), (f) linearized $CO_2$sorption kinetics plotted as ln $(q^*_{CO_2} - q_{CO_2})$ vs. time; (c), (g) $H_2O$ cumulative loading with corresponding (d), (h) linearized $H_2O$ sorption kinetics plotted as $ln(q^*_{H_2O} - q_{H_2O})$ vs. time. Dotted lines represent the linear fit of the third cycle.

Mass transfer coefficients ($k$) for both $CO_2$ and $H_2O$ were estimated from the slope of the linearized time-dependent loading data according to eqn (7). For each cycle, $q_i(0)$ in eqn (7) was taken as zero and $q_i^*$ was calculated as the average loading of the last 60 minutes to account for the fluctuations in data once steady state was approached. For both 95% → 20% RH and 20% → 95% RH cycles, $k_{CO_2}$ was obtained from the approximately linear region of the first 100 minutes. The goodness of fit for the mass transfer coefficients of $CO_2$ and $H_2O$ are shown as dotted lines using cycle 3 data as shown in Fig. 5(b, d, f, h). Some deviation from linearity is evident in nearly all cases, suggesting more complex kinetic behavior. Nevertheless, the linear driving force approximation was retained because a more detailed mechanistic kinetic model is not yet available.

The average of the three cycles was used to represent the mass transfer coefficients in the VMS simulations. The resulting average $k_{CO_2}$ values were $5.1 \pm 0.6 \text{ x } 10^{-4} s^{-1}$ for $CO_2$ sorption (95% → 20% RH) and $4.0 \pm 0.6 \text{ x } 10^{-4} s^{-1}$ for $CO_2$ desorption (20% → 95% RH), where the error represents standard deviation. These results indicate that $CO_2$ sorption during drying is approximately 25% faster than desorption, despite exhibiting an initial non-linear lag during the first 10-20 minutes. This lag in $CO_2$ uptake likely reflects the initial desorption of water, which must occur before $CO_2$ can bind to the active sites. The increase in kinetics of sorption vs. desorption is possibly due to faster transport within a hydrated vs dehydrated polymer. Similar results were observed for IRA900 in recent dynamic breakthrough column studies.[36] Amine based sorption models employ $CO_2$ mass transfer coefficients ranging from 2-50 x $10^{-4}$ $s^{-1}$.[23,29,31,33] The lower-end values, which are comparable to those observed for IRA900, originate from $CO_2$ uptake measurements of similar design and timescale.[23,29]

For $H_2O$, $k_{H_2O}$ was fit to the linear portion of the uptake and release curves data using the first 45 minutes for the 95% → 20% RH transition and the first 50 minutes for 20% → 95% RH. Beyond these windows, fluctuations in steady state loading introduced significant non-linearity that compromised the linearized approximation. The resulting average mass transfer coefficients were $14 \pm 4 \text{ x } 10^{-4} s^{-1}$ during 95% → 20% RH and $4.5 \pm 0.2 \text{ x } 10^{-4} s^{-1}$ during 20% → 95% RH. These results indicate that water sorption during the vapor stripping step is nearly three times slower than water desorption, a trend also reflected in the cumulative loading curves in Fig. 5(c) & Fig. 5(g). The slower desorption kinetics are comparable to those of $CO_2$ (de)sorption, suggesting that both species experience similar intraparticle kinetic constraints under these conditions.

Notably, the mass-transfer coefficients observed here are nearly an order of magnitude slower than those assumed in comparable sorption models ($20 - 90 \text{ x } 10^{-4} s^{-1}$ ).[23,29,31] The slower kinetics of $CO_2$ and $H_2O$ likely arise from intraparticle transport limitations, like sorbate diffusion into a charged, glassy polymer matrix, rather than vapor diffusion in macropores. The asymmetry between $H_2O$ sorption and desorption may stem from concentration-dependent changes in water diffusivity in the polymer matrix.[54,55] Detailed kinetic studies are likely necessary to understand the ultimate transport limitations to design sorbent structures that accommodate both high $CO_2$ capacity and fast transport.

## 3.2 VMS behavior

**Verification to Haghpanah PSA $CO_2$**

To validate the accuracy of the numerical framework, the numerical model adapted in this study was compared against an established pressure vacuum swing adsorption (PVSA) and associated bed geometry described in Section 2.2.4. The transient behavior of a $CO_2$ adsorption step in a packed bed of zeolite sorbent provides a benchmark for evaluating coupled heat-, momentum- and mass-transfer predictions. Fig. 6 shows the resulting transient behavior of dimensionless outlet velocity$(v/v_0)$, temperature $(T/T_0)$ and $CO_2$ mole fraction $(y_{CO_2})$. Two distinct transitions are evident. The first, at approximately 500 s, corresponds to $CO_2$ breakthrough which is marked by a sharp temperature rise due to the exothermic nature of sorption and an accompanying increase in outlet velocity due to gas expansion. The second transition near 3600 s, reflects completion of $CO_2$ adsorption as the mole fraction increases to the level of the inlet, with subsequent thermal elution and velocity decline as the cooled gas contracts.

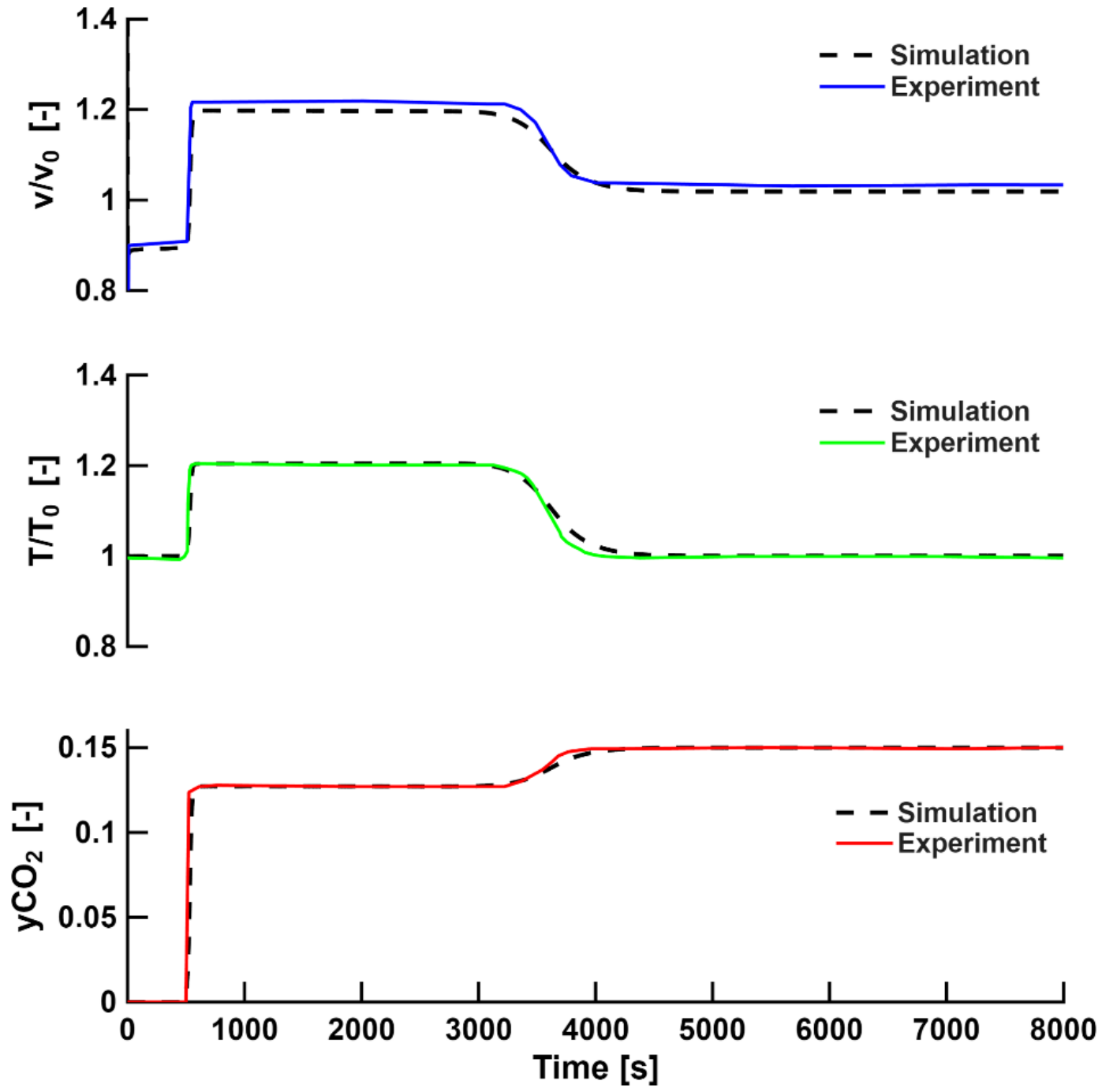

Fig. 6 Comparison of dimensionless velocity, temperature, and $CO_2$ mole fraction breakthrough profiles at the outlet of the adsorption column for a PSA system with zeolite sorbent. Solid colored lines denote the simulated results reported by Haghpanah et al.,[48] validated against their experimental data while the dotted black lines represent the predictions obtained in this study.

In each subplot in Fig. 6, the solid colored curves represent the experimentally verified results from the reference study,[48] while the dashed black lines correspond to the present numerical model's predictions. The close agreement demonstrates that the implemented governing equations, kinetic equations and thermodynamic parameters of the matching zeolite material collectively capture the physics governing an adsorption process. Further, mass and energy balance verifications were performed, with the corresponding equations provided in SI eqn (1-4) for mass balance and eqn (5-11) for energy balance. The resulting errors were well within the margin reported in the reference literature —2% for the mass balance error and less than 0.4% for the energy balance—confirming mass conservation and numerical stability.

**Spatially averaged temporal response**

Building upon this verified framework, the model was adapted to simulate the VMS cycle using the moisture-swing sorbent thermodynamics (see section 2.1) and kinetics (see section 3.1) together with a vacuum-vapor-stripping regeneration mode. A CSS case study was conducted to investigate the behavior of the VMS process, including $CO_2$ and $H_2O$ loading, pressure and temperature responses, and the determination of step durations based on each step's operational objective. Baseline operating conditions are summarized in **Table 4**.

Air and vapor stripping velocity were both set to 1 $m \cdot s^{-1}$ respectively, reflecting optimized air velocities found in other work.[16,23] The intermediate pressure ($P_I$) of the air evacuation step is 0.5 kPa, ensuring $CO_2$ purity >99% and enabling boiling of water for the subsequent vapor stripping step. After vapor stripping, the final desorption column pressure ($P_L$) is set to 1 kPa to pull residual $CO_2$ and water vapor out of the bed. The feed air contacting the column during $CO_2$ sorption is assumed to be 298K and 20% RH to align with the sorbent kinetic measurements, and temperature of the water reservoir that dictates the water saturation pressure ($P_{sat}$) is set to 300K, based on arid region seawater or brackish groundwater temperatures. Water vapor saturation pressures were assumed based on seawater corrections.[56]

Pressurization and evacuation times were brief transition steps used to achieve target pressure and were set to approximately 10 s for pressurization to reach 100 kPa and 30 s for air evacuation to reach 0.5 kPa. The duration of the $CO_2$ sorption and vapor stripping steps were intentionally held long (200 and 400 min., respectively) to observe the effect of full loading and unloading of $CO_2$.

**Table 4 Baseline operating conditions for the VMS case study**

| | |
|---|---|
| **Pressurization** | |
| atmospheric pressure $P_{atm}$ [kPa] | 100 |
| air temperature, $T_{AIR}$ [k] | 298 |
| duration, $t_{PZ}$, [s] | 10 |
| **$CO_2$ Sorption** | |
| feed air velocity, $v_{AIR}$ [m/s] | 1 |
| air temperature, $T_{AIR}$ [k] | 298 |
| feed air relative humidity, $RH_{AIR}$ [%] | 20 |
| duration, $t_{CS}$, [s] | 12000 |
| **Air Evacuation** | |
| intermediate pressure, $P_I$ [kPa] | 0.5 |
| duration, $t_{EV}$, [s] | 30 |
| **Vapor Stripping** | |
| vapor stripping velocity, $v_{VS}$ [m/s] | 1 |
| water saturation temperature, $T_{sat}$ [k] | 300 |
| vapor relative humidity, $RH_{VS}$ [%] | 100 |
| duration, $t_{VS}$, [s] | 24000 |
| **Final Desorption** | |
| low pressure, $P_L$, [kPa] | 1 |
| duration, $t_{DES}$, [s] | 200 |

Fig. 7(a-d) presents the CSS cycle of spatially averaged bed properties: (a) $CO_2$ loading, (b) $H_2O$ loading, (c) total column pressure and (d) total column temperature. Cycle steps are shown in color: pressurization (red), $CO_2$ sorption (green), air evacuation (blue), vapor stripping (purple) and final desorption (burgundy). During $CO_2$ sorption (Fig. 7(a)), most uptake occurs within the first 100 min, after which the rate slows as the sorbent nears equilibrium. Only minor $CO_2$ uptake occurs during the preceding pressurization step because the sorbent remains heavily loaded with $H_2O$ from a prior vapor stripping step. Similarly, water loading remains largely unchanged during pressurization. The corresponding water loading (Fig. 7(b)) during $CO_2$ sorption shows a sharp decrease at the beginning as it evaporates, followed by a gradual decline when approaching equilibrium with the ambient air. The spatially averaged column pressure remains at 100 kPa throughout this step (Fig. 7(c)), while spatially-averaged temperature increases to ambient conditions (Fig. 7(d)). The short duration of air evacuation (blue) does little to change the $CO_2$ and $H_2O$ loading of the sorbent, but the vacuum pumping extracts excess air in the sorbent bed dropping the pressure to 0.5 kPa, producing a small but noticeable cooling effect (~1.5K).

During vapor stripping (purple), water vapor is introduced to drive $CO_2$ release, consistent with the moisture swing behavior described in eqn (1). It takes roughly twice as long to release than bind $CO_2$. The longer desorption time reflects the slower kinetics described in section 3.1, where desorption proceeds more slowly than $CO_2$ sorption and the slower kinetics of the water sorption. Relative to the rate of $H_2O$ evaporation during the $CO_2$ sorption step, $H_2O$ loading in vapor stripping occurs more slowly because the mass transfer coefficient $\left(k_{\mathrm{H_2O}}^{ads}\right)$ during vapor stripping is three times lower than $\left(k_{\mathrm{H_2O}}^{des}\right)$. Said differently, the kinetics of $CO_2$ desorption are limited by rates of $H_2O$ sorption, as water loading directly affects the $CO_2$ bi-component isotherm (eqn (5)). This behavior contrasts with the TVSA amine-based systems, where $CO_2$ uptake typically requires longer residence times than high-temperature regeneration. Finally, a spike in temperature is observed during the vapor stripping step (Fig. 7(d)) which reflects the exothermic nature of water sorption. Temperature slowly decreases as further water loading (and the associated exotherm) slows, and thermal energy is advected out with the desorbed $CO_2$ and water vapor. A short final desorption step (red) was sustained for 200 s to remove residual $CO_2$ and $H_2O$ driving pressure below water vapor saturation pressures (1 kPa). This final purge produces a small decrease in $H_2O$ loading and an associated drop in bed temperature, resetting conditions for the subsequent cycle.

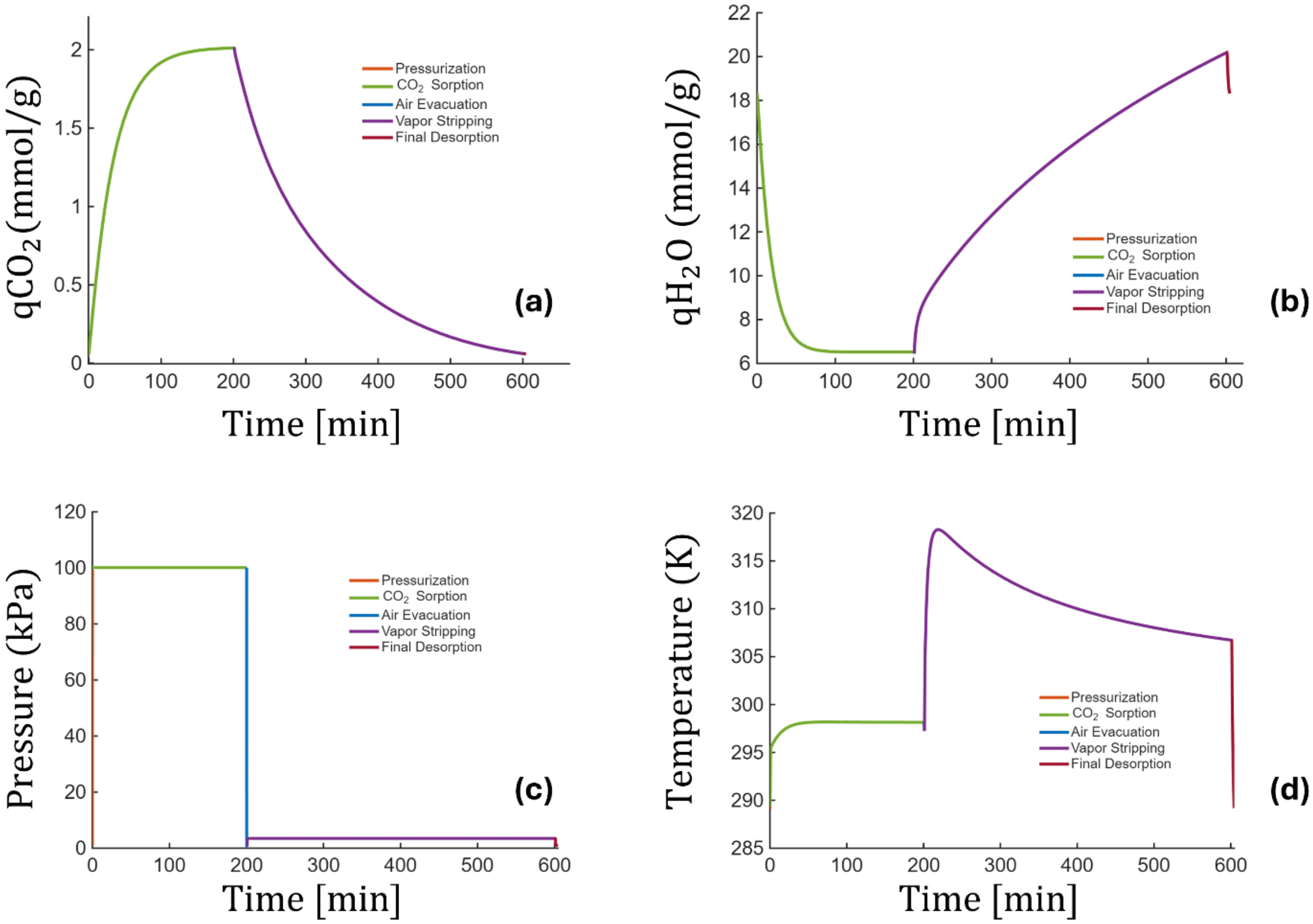

Fig. 7 Temporal profiles of spatially averaged variables—a) $CO_2$ loading, b) $H_2O$ loading, c) total pressure, and d) temperature—during the pressurization, $CO_2$ sorption, air evacuation, vapor stripping, and desorption steps for the case study. The step durations are pressurization (10 s), $CO_2$ sorption (200 min), air evacuation (30 s), vapor stripping (400 min), and desorption (200 s).

Since desorption in the VMS process relies solely on vacuum and vapor without external heating, the desorption rate is fundamentally limited by the low vapor pressure at near-ambient temperatures. Although not considered in this work, the thermal behavior in Fig. 7(d) suggests a potential pathway to harvest the exothermic heat released during water sorption via condenser heat exchange with the feed water reservoir to increase the vapor pressure in the vapor stripping step. This elevated vapor pressure (and temperature) may accelerate sorption kinetics while also reducing the pressure ratio energy demand from the vacuum blower (eqn. 24), in turn improving the productivity and lowering the energy penalty while preserving the low-temperature operational advantages of the VMS process when compared to steam assisted TVSA systems.[23,29,30]

## 3.3 VMS optimization

*Energy, Productivity & Water Loss*

To fully interpret the productivity, energy and water utilization of the VMS process, the dynamic sorption behaviors captured in Fig. 7 provide a quantitative basis for establishing realistic operating bounds and identifying operating variables that strongly influence performance, which are used for optimization. Fig.8 presents the resulting optimized tradeoff between the objectives under 20% RH ambient conditions. The robustness of this optimization simulation was verified through multiple runs with varied initial population sampling and initial column properties, all of which yielded consistent Pareto-optimal trends (Fig S1).

Analysis of the behavior in Fig. 8 reveals a clear progression in system performance as productivity is increased. At low productivity, the system operates with minimal electrical energy consumption and low water loss. As productivity increases, both electrical energy consumption and water loss increase, with water loss exhibiting an approximately linear dependence on productivity, while the energy requirement rises modestly at first and then very sharply as higher-throughput is pursued. This transition marks the onset of rapidly diminishing returns. In particular, beyond a productivity of 0.52 $\mathrm{kgCO_2 \cdot kg_s^{-1} \cdot d^{-1}}$, achieving a ~15% increase toward the maximum optimized productivity of 0.61 $\mathrm{kgCO_2 \cdot kg_s^{-1} \cdot d^{-1}}$ requires disproportionately large increases in energy from ~4.5 to ~17 $\mathrm{MJ \cdot kgCO_2}^{-1}$ along with higher water losses.

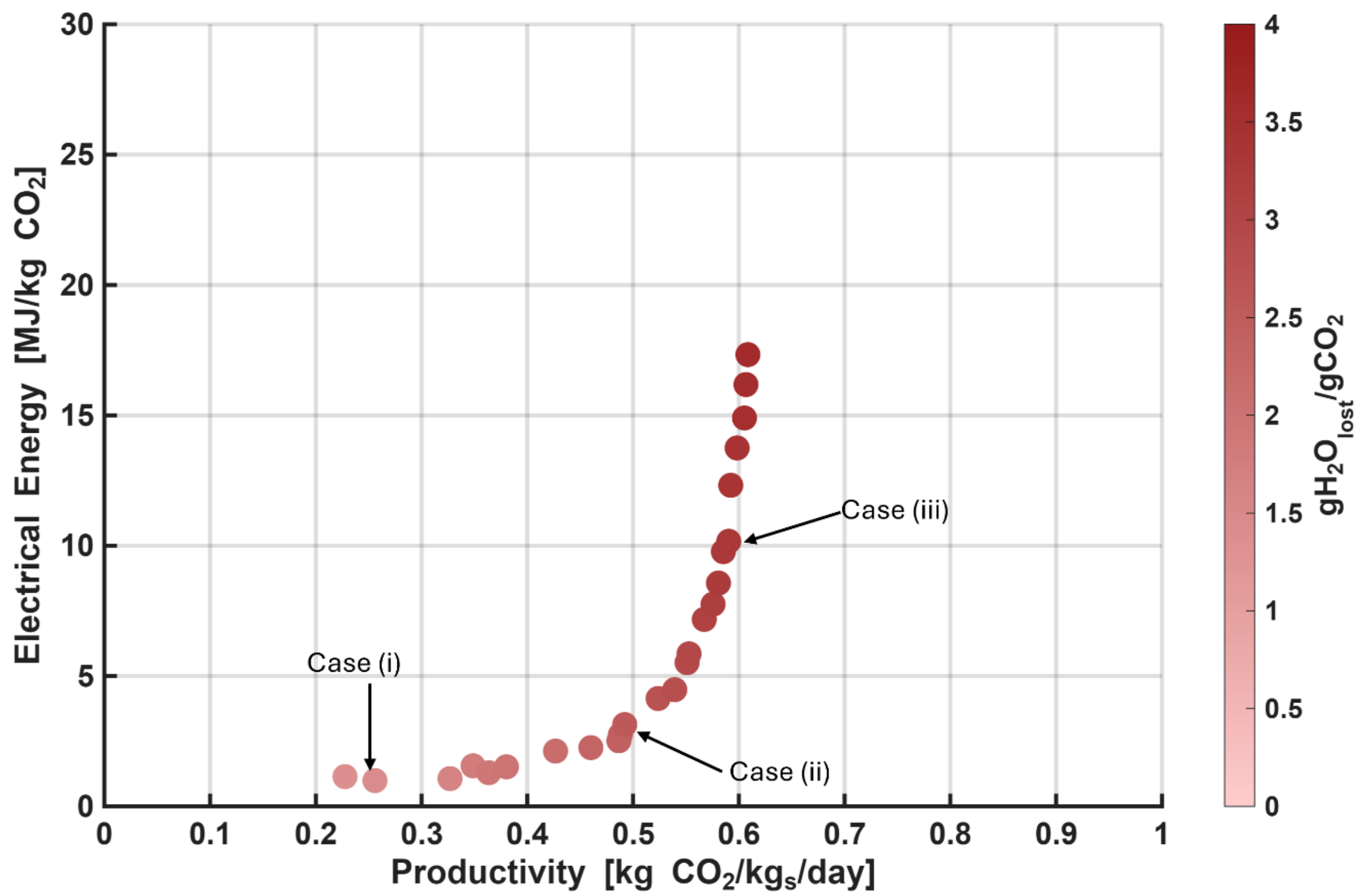

Fig. 8 Pareto front for the maximization of productivity, minimization of both electrical and water loss at 20% RH feed ambient conditions.

These observations indicate that the tradeoffs among productivity, energy consumption, and water loss are governed by a sensitive interplay of operating variables, motivating a more detailed examination of how the decision variables shape these outcomes. Fig. 9 presents the optimization objectives as a function of each decision variable providing a global view of how performance varies across the optimal design space. Each point represents a distinct optimized combination of velocities and step durations and therefore must be interpreted in the context of the other decision variables rather than as a one-at-a-time parametric trend. Low energy and minimal water loss operating points are primarily associated with slower $v_{\mathrm{AIR}}$ (Fig. 9 a,i) and $v_{VS}$ (Fig. 9 b,j) and correspondingly longer step durations (Fig. 9(c,d,k,l)), conditions that yield low $CO_2$ productivity. Higher velocities strengthen convective mass transfer and simultaneously depress the surrounding gas-phase partial pressures of water vapor (in the $CO_2$ sorption step) or $CO_2$ (in the stripping step), collectively increasing the driving force and rate of reversible sorption for both sorbates. However, operation beyond $v_{VS}$= ~1.7 m·s$^{-1}$, only yields marginal gains in $CO_2$ throughput (Fig. 9f). This is evidenced by the fact that, over the $v_{VS}$ range of 1.7-5 m·s$^{-1}$ with $v_{AIR}$ between 0.8-1 m·s$^{-1}$, the optimized solutions maintain similarly short step durations. This indicates that the system is near saturation-limited conditions, so additional convective driving force offers little benefit. This operating regime reflects the transition observed in Fig. 8 where the electrical energy consumption increases sharply, highlighting that aggressive transport intensification primarily penalizes energy and water usage rather than improving performance.

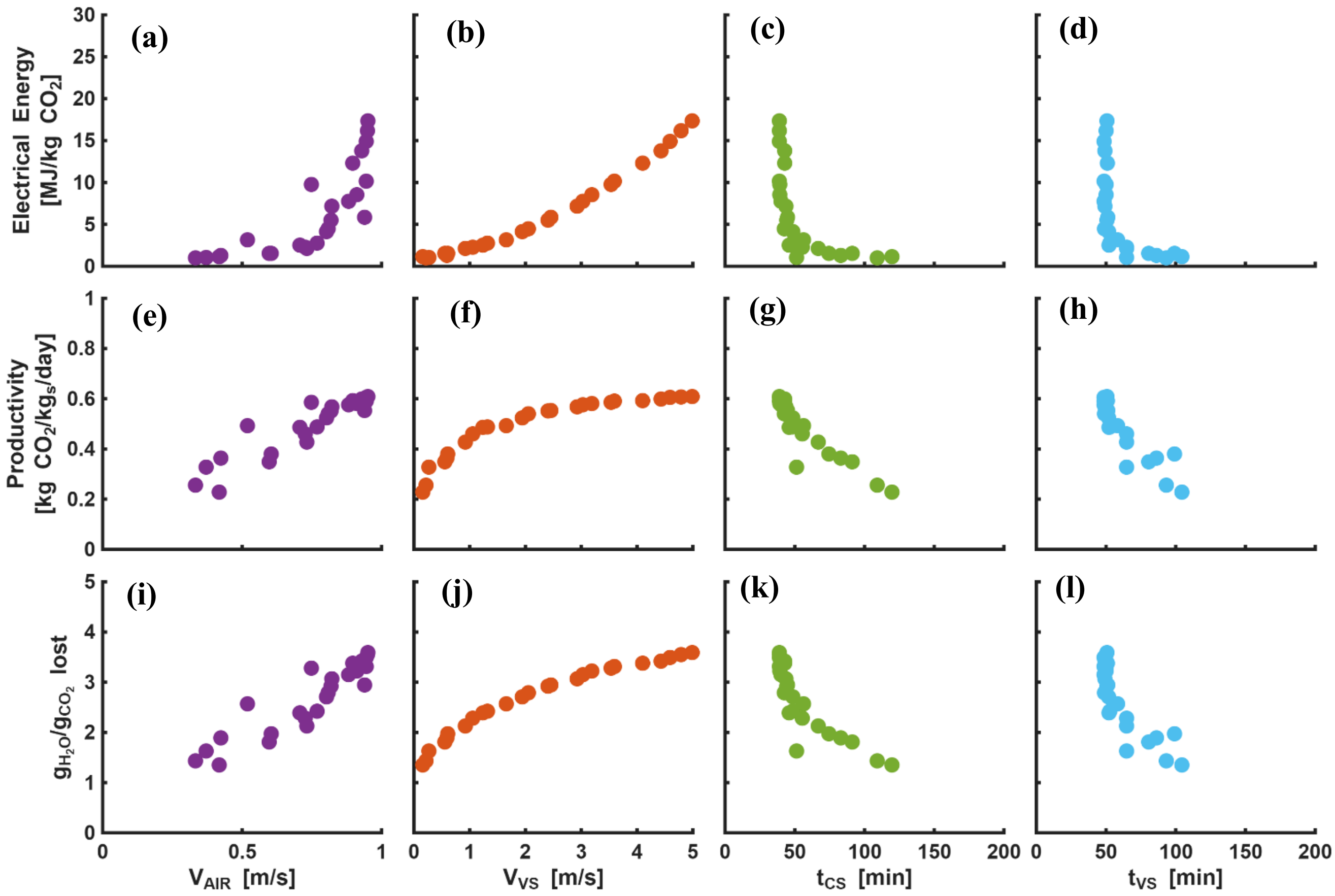

Fig. 9 Electrical energy, $CO_2$ productivity and water loss presented as a function of (a), (e), (i) air velocity, (b), (f), (j) vapor stripping velocity, (c), (g), (k) $CO_2$ sorption time & (d), (h), (l) vapor stripping time respectively. Air feed = 20% RH and 25 °C, water supply is isothermal at $T_{sat}$ = 27 °C with $P_{sat} = 3.5kPa$.

Moving beyond these global trends, we identify the unit operations responsible for the rising energy and water penalties. To achieve this, we select three optimized cases spanning different velocities and durations (notated in Fig. 8 and summarized in Table 5) and present their stepwise contributions to the total specific energy in Fig.10. The three cases pair faster velocities with shorter durations and vice versa to balance the competing objectives. In case (i) where $v_{AIR} > v_{VS}$, $CO_2$ sorption accounts for 22% ($0.21\ \mathrm{MJ} \cdot \mathrm{kgCO_2}^{-1}$) of the total energy while vapor stripping contributes only 7% ($0.07\ \mathrm{MJ} \cdot \mathrm{kgCO_2}^{-1}$). In this case, $CO_2$ compression dominates the energy demand (~$0.47\ \mathrm{MJ} \cdot \mathrm{kgCO_2}^{-1}$), driven by compression of the captured $CO_2$ to $P_{atm}$ (eqn. 28), and remains nearly constant across all three cases.

**Table 5. Three representative Optimized cases for the electrical energy analysis**

| Case | $v_{AIR}$ [m/s] | $v_{VS}$ [m/s] | $t_{CS}$ [min] | $t_{VS}$ [min] | $t_{DES}$ [min] | $E_T$ [MJ /kgCO$_2$] | Pr [kgCO$_2 \cdot kg_s^{-1} \cdot d^{-1}$] | $H_2O_{loss}$ [gH$_2$O /gCO$_2$] |
|---|---|---|---|---|---|---|---|---|
| (i) | 0.33 | 0.22 | 109 | 93 | 3.7 | 0.94 | 0.26 | 1.4 |
| (ii) | 0.71 | 1.2 | 46 | 52 | 3.8 | 2.50 | 0.50 | 2.4 |
| (iii) | 0.75 | 3.5 | 39 | 50 | 3.9 | 9.50 | 0.60 | 3.3 |

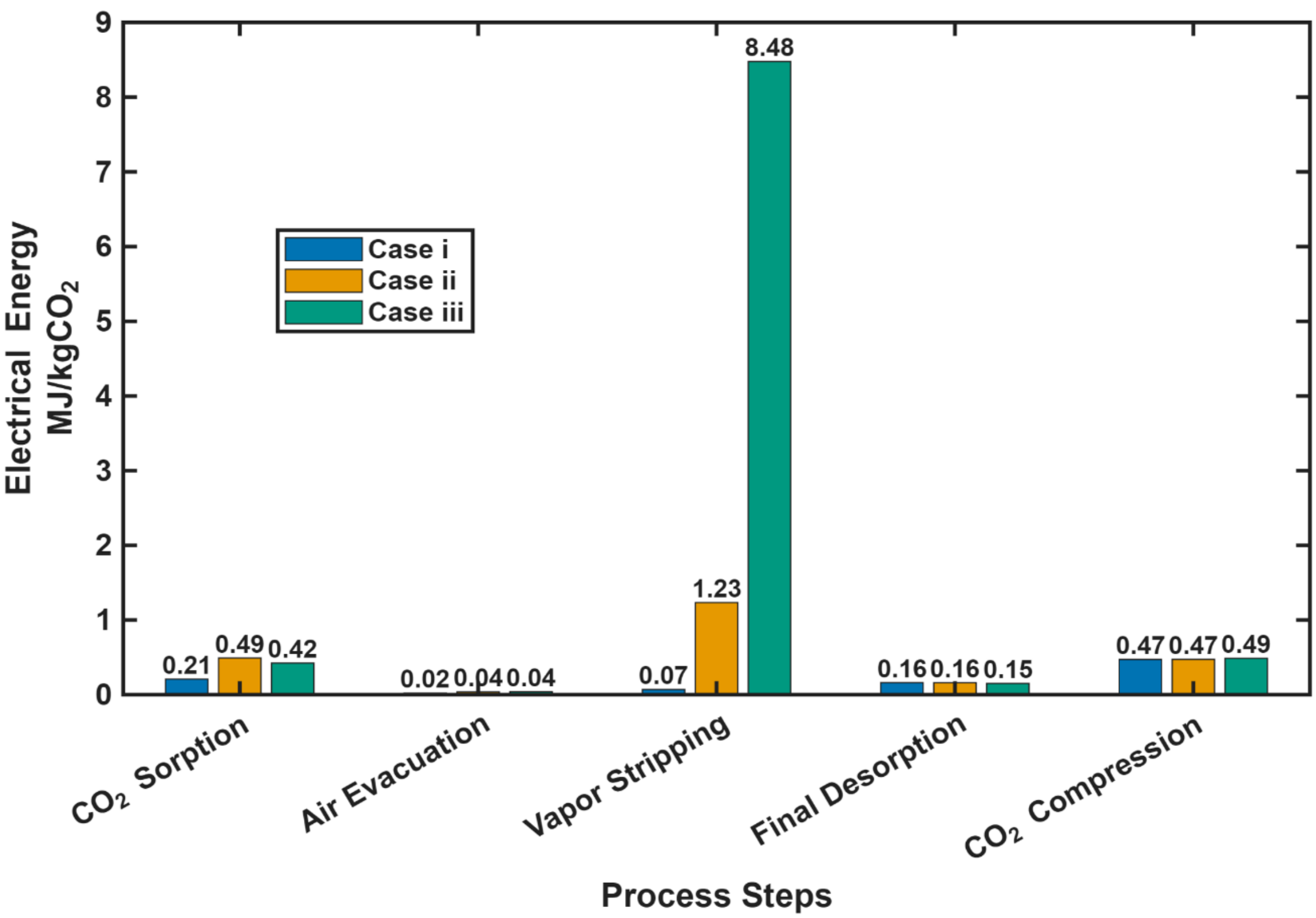

Fig. 10 Stepwise specific energy decomposition for the three representative optimized cases, showing the contributions from $CO_2$ sorption, air evacuation, vapor stripping, final desorption, and $CO_2$ compression. Vapor stripping and final desorption involve processing only water vapor and compression to $P_{sat}$ respectively, whereas $CO_2$ compression corresponds exclusively to compressing the recovered $CO_2$ stream to $P_{atm}$.

In contrast, vapor stripping becomes the largest energy consumer in cases (ii) and (iii), where $v_{AIR} < v_{VS}$, accounting for 49% ($1.2\ MJ \cdot kgCO_2{}^{-1}$) and 89% ($8.5\ MJ \cdot kgCO_2{}^{-1}$) of the total energy respectively. This shift results from the higher vapor velocities, which substantially increase the water molar flow rate and thus turbomachinery work, despite the lower compression ratio relative to $CO_2$ (eqn. 26). Although $CO_2$ sorption contributes less energy than vapor stripping in cases (ii) and (iii), this reflects the lower operating velocities in the sorption step and would be expected to rise sharply at higher air velocities due to the growing importance of the kinetic term in the Ergun pressure-drop expression.

Meanwhile, air evacuation remains a minor contributor despite applied vacuum below saturated pressures, owing to its short step duration. Final desorption likewise remains nearly constant (0.16 $\mathrm{MJ \cdot kgCO_2}^{-1}$) because its duration is consistently ~4 min and involves compression only to $P_{sat}$. Consequently, its contribution to the objective sensitivity is negligible and it is therefore omitted from Fig. 9, despite being a decision variable.

Notably, water losses increase systematically from case (i) - case (iii), indicating that more aggressive vapor processing leads to greater water uptake by the sorbent, which is ultimately lost during $CO_2$ sorption. This trend is illustrated in Fig. 9(i), where at a fixed $v_{AIR}$ of approximately 0.75 m·s$^{-1}$ two distinct water loss values are observed, ~2.1 and ~ 3.3 $\mathrm{gH_2O \cdot gCO_2}^{-1}$, corresponding to an increase in $v_{VS}$ from ~0.9 to ~3.5 m·s$^{-1}$. This directly confirms that increasing the amount of vapor processed leads to increased water losses, even at the same air velocity.

A more detailed view of how air and vapor velocities independently affect the VMS process is provided in the supporting information (Fig. S2). When $v_{AIR}$ is modulated between 0.1 – 3 m·s$^{-1}$ at a fixed $v_{VS}$ = 1 m·s$^{-1}$, $CO_2$ loading increases with air velocity up to ~1 m·s$^{-1}$, beyond which no further gains occur (Fig. S2(a)). In contrast, varying $v_{VS}$ from 0.1 – 3 m·s$^{-1}$ at fixed $v_{air}$ = 1 m·s$^{-1}$, shows that $CO_2$ desorption continues to accelerate with increasing vapor velocity (Fig. S2(b)). Water loading follows a similar pattern: higher air velocities beyond 1 m/s does little to increase rates of $H_2O$ desorption in the $CO_2$ sorption step, the ultimate cause of evaporative water loss (Fig. S2(b)), whereas increasing vapor velocity strongly increases water uptake and thus the amount of water that can be subsequently lost during $CO_2$ sorption (Fig. S2(e)).

Velocity also influences sorbent bed temperature. Slow air velocities induce sub-ambient cooling due to strong water evaporative demand (Eqn. 12), whereas higher air velocities more rapidly equilibrate the bed with ambient conditions (Fig. S2(c). Conversely, slow vapor velocities produce larger autothermal temperature rises during desorption because the heat of water sorption dominates over convective heat removal. Finally, the depth of the final desorption vacuum further affects temperature: reducing the bed pressure from near water saturation (3–5 kPa) to 1 kPa yields an additional ~15 K temperature drop.

Taken together, the optimization results show that no single step or operating variable is intrinsically dominant; rather, total specific energy is governed by the coupled effects of velocity-dependent pressure drop and molar flow rate, step duration and $CO_2$ compression level, and the resulting $CO_2$ productivity on which energy is normalized. Water losses are minimized by processing only enough vapor to achieve near-complete $CO_2$ desorption, while productivity is maximized by balancing the sorption and stripping steps to avoid both sorbent underutilization and excessive water throughput with diminishing returns.

Consistent with these trends, the 20% RH ambient air optimization identifies a practical operating window delivering competitive performance at productivities of ~0.23–0.50 $\mathrm{kgCO_2 \cdot kg_s^{-1} \cdot d^{-1}}$, with specific energies of ~1–2.5 $\mathrm{MJ \cdot kgCO_2}^{-1}$, and water losses of ~1.3–2.4 $\mathrm{gH_2O \cdot gCO_2}^{-1}$.

These metrics are achieved for $v_{\text{Air}} \approx 0.3–0.7\ \text{m}\cdot\text{s}^{-1}$, $v_{\text{VS}} \approx 0.2–1.2\ \text{m}\cdot\text{s}^{-1}$, 46 min < $t_{\text{CS}}$ < 120 min, and 52 min< $t_{\text{VS}}$ < 104 min.

Although not an optimized target, $CO_2$ purity was tracked and remained consistently high (~99%) across all solutions. This high purity arises from the vacuum pressure (0.5 kPa) used during the air evacuation, step 3, to remove residual air. While a detailed sensitivity analysis was not performed, $CO_2$ purity is expected to decline with more moderate evacuation pressures due to greater retention of residual air. Steps 4 and 5 further assume full separation of $CO_2$ from $H_2O$ as $CO_2$ is compressed above water vapor saturation pressures.

## 3.4 VMS parameterization

### *3.4.1 Relative Humidity*

Because the properties of the inlet air depend strongly on ambient weather conditions, particularly its humidity and temperature, VMS performance is expected to vary with these environmental factors. Here we evaluate the impact of air relative humidity (RH), as this parameter has a pronounced influence on the moisture-swing activity of the sorbent. To quantify this effect, additional optimization studies were conducted at 50%, 65%, and 80% RH for comparison against the baseline 20% RH air. **Table 6** summarizes the operating conditions that correspond to minimum energy, maximum productivity, and minimum water loss—reflecting the three optimization objectives at each humidity level, while Fig.11 depicts the overall optimized landscape.

**Table 6 Optimized operating conditions corresponding to minimum energy, maximum productivity and minimum water loss at three different relative humidities, 20%, 50%, 65% and 80%.**

| RH | Objective | $t_{\text{CS}}$ [min] | $t_{\text{VS}}$ [min] | $t_{\text{DES}}$ [min] | $v_{\text{AIR}}$ [m/s] | $v_{\text{VS}}$ [m/s] | $E_T$ [MJ/ kg $CO_2$] | $Pr$ [kg$CO_2$ $\text{kg}_s^{-1}$ $\text{day}^{-1}$] | $H_2O$ Loss [g$H_2O$/ g$CO_2$] |
|---|---|---|---|---|---|---|---|---|---|
| 20% | (Min $E_T$) | 109 | 93 | 3.7 | 0.33 | 0.22 | 1.0 | 0.26 | 1.4 |
| | (Max Pr) | 39 | 51 | 3.9 | 0.95 | 5.00 | 15 | 0.61 | 3.6 |
| | (Min $H_2O$ loss) | 120 | 104 | 3.7 | 0.42 | 0.16 | 1.1 | 0.23 | 1.4 |
| 50% | (Min $E_T$) | 116 | 173 | 5.9 | 0.34 | 0.32 | 1.4 | 0.25 | 0.9 |
| | (Max Pr) | 40 | 47 | 4.5 | 0.97 | 4.80 | 17 | 0.56 | 2.0 |
| | (Min $H_2O$ loss) | 64 | 52 | 4.7 | 0.76 | 0.17 | 3.4 | 0.22 | 0.31 |
| 65% | (Min $E_T$) | 111 | 138 | 4.3 | 0.26 | 0.35 | 1.8 | 0.24 | 0.5 |
| | (Max Pr) | 36 | 39 | 6.0 | 1.0 | 4.83 | 20 | 0.48 | 1.1 |
| | (Min $H_2O$ loss) | 82 | 57 | 4.4 | 0.90 | 0.28 | 4.2 | 0.24 | $2x10^{-2}$ |
| 80% | (Min $E_T$) | 139 | 113 | 4.1 | 0.36 | 0.82 | 3.4 | 0.22 | $2x10^{-1}$ |
| | (Max Pr) | 74 | 57 | 4.1 | 0.94 | 4.96 | 28 | 0.32 | $3x10^{-2}$ |
| | (Min $H_2O$ loss) | 90 | 67 | 4.2 | 0.74 | 1.22 | 5.3 | 0.28 | $8x10^{-4}$ |

Fig. 11 illustrates how optimized VMS performance responds to ambient air relative humidity. Lower inlet RH environments enable higher productivity at lower specific energy demand, but inherently result in greater evaporative water loss (darker color). As ambient RH increases, optimized productivity and energy performance degrades, while water loss is progressively reduced. This sensitivity to humidity becomes most pronounced under very humid conditions (80%), where productivity is strongly suppressed despite negligible water loss. Evaluated at a representative specific energy demand of $5\ \mathrm{MJ} \cdot \mathrm{kgCO_2}^{-1}$, operation under 20% RH achieves the highest productivity ($0.55\ \mathrm{kgCO_2} \cdot \mathrm{kg_s^{-1}} \cdot \mathrm{d^{-1}}$), but also incurs the largest water loss of $2.9\ \mathrm{gH_2O} \cdot \mathrm{gCO_2}^{-1}$. At 50% RH, productivity decreases by 10% to $0.5\ \mathrm{kgCO_2} \cdot \mathrm{kg_s^{-1}} \cdot \mathrm{d^{-1}}$ while water loss is reduced by roughly half to $1.4\ \mathrm{gH_2O} \cdot \mathrm{gCO_2}^{-1}$. Increasing ambient humidity to 65% further lowers productivity by a similar margin to $0.44\ \mathrm{kgCO_2} \cdot \mathrm{kg_s^{-1}} \cdot \mathrm{d^{-1}}$, accompanied by a substantial drop in water loss to $0.65\ \mathrm{gH_2O} \cdot \mathrm{gCO_2}^{-1}$. Under very humid conditions (80% RH) productivity drops sharply to $0.28\ \mathrm{kgCO_2} \cdot \mathrm{kg_s^{-1}} \cdot \mathrm{d^{-1}}$, with near-zero net water loss. At RH approaching saturation, productivity approaches zero and operation is no longer viable. Collectively, these results demonstrate that VMS remains operable across a wide range of ambient humidities, while indicating that moderate humidity conditions, despite modest productivity penalties, offer substantial water savings that are likely to influence optimal operation based on local water availability and cost. Diurnal variations in ambient RH (and associated air temperature) are therefore expected to drive an inherently dynamic VMS operation.

This humidity dependence arises from the moisture-swing $CO_2$ isotherm, in which the negative power-law exponent *n* applied to the water-loading-dependent affinity term, $K_{MS}[q_{H_2O}]^{-n}$(eqn. 5), induces a nonlinear collapse in $CO_2$ affinity with increasing water uptake. Because water loading itself increases nonlinearly with ambient RH (Fig. 2), elevated humidity disproportionately suppresses $CO_2$ adsorption. Under dry conditions, low water coverage preserves a high effective affinity and enables strong $CO_2$ uptake, whereas progressively more humid air increases water activity and weakens sorption. Consequently, performance at 50% RH remains much closer to that at 20% RH than at 80% RH, where increased water activity drives a sharp loss in $CO_2$ capacity, consistent with the isotherm curvature observed in Fig. 2a.

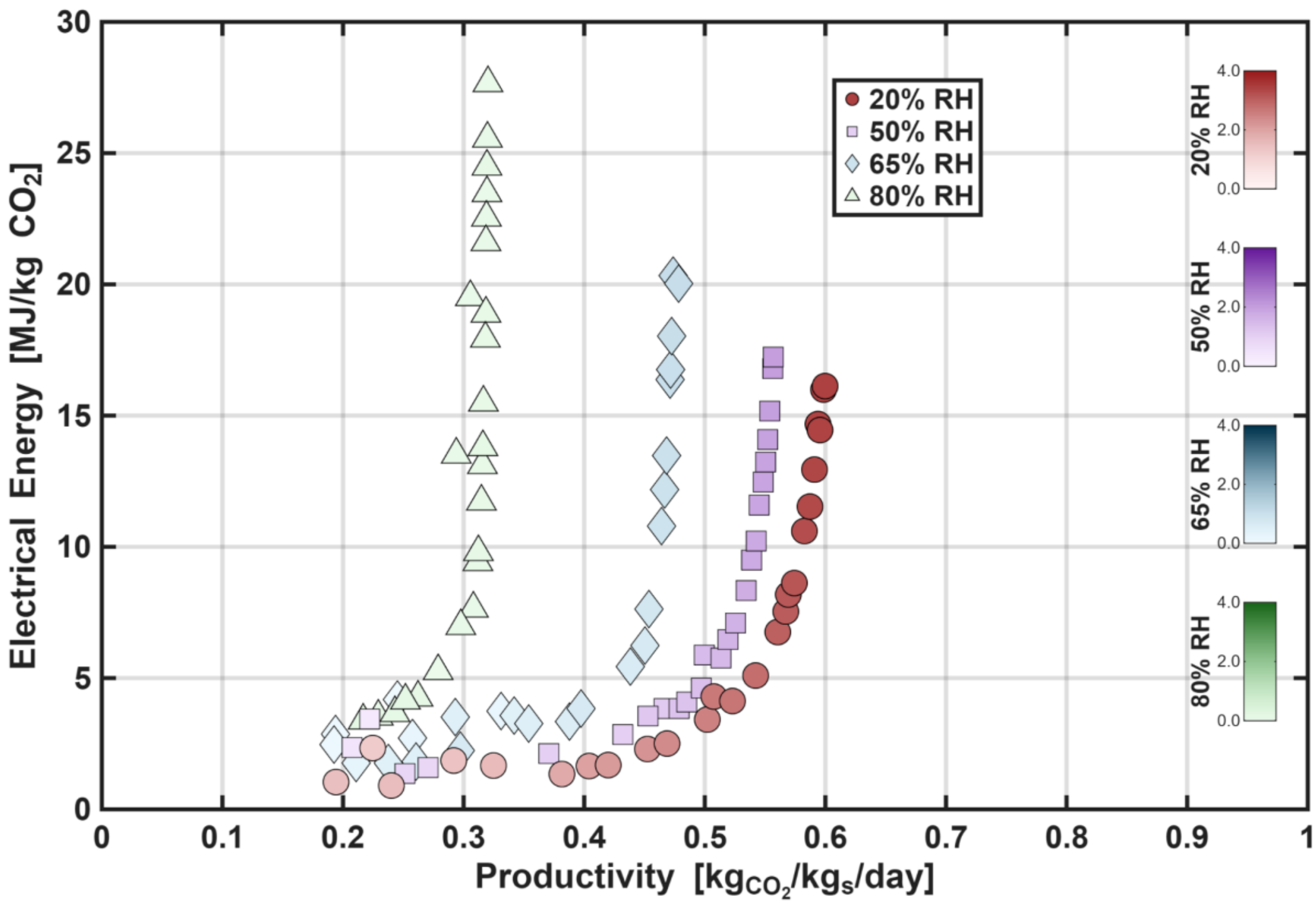

Fig. 11 Non-dominated optimized solutions show the relationship between Electrical Energy $(MJ \cdot kg_{CO2}{}^{-1})$ and productivity $(kg\ CO_2/kg_s/day)$ with the color density representing water loss gradient across three different inlet air humidities—20% (burgundy), 50% (purple) , 65% (blue) and 80% (green)

### *3.4.2 Kinetics*

Because sorbent utilization is kinetically limited, mass transport rates are a key driver of $CO_2$ productivity. Yet their influence on water loss is less clear, as changes in mass transfer can alter water loss in non-obvious ways. To understand this effect, we parameterize the process across a range of kinetic regimes. In Fig. 12 122, we evaluate three kinetic scenarios — slow (half the baseline), baseline (measured kinetics from section 3.1) and fast (2x the baseline)—all under 20% RH feed air. For each case, the mass transfer coefficients for $CO_2$ and $H_2O$ in both the sorption and vapor stripping steps are varied, as summarized in **Table 7**. The results are presented from the perspective of the trade-off between $CO_2$ capture productivity and water loss, revealing a clear Pareto front. The color bar indicates the corresponding energy penalty, spanning 0.8-30 $\mathrm{MJ} \cdot \mathrm{kgCO_2}^{-1}$.

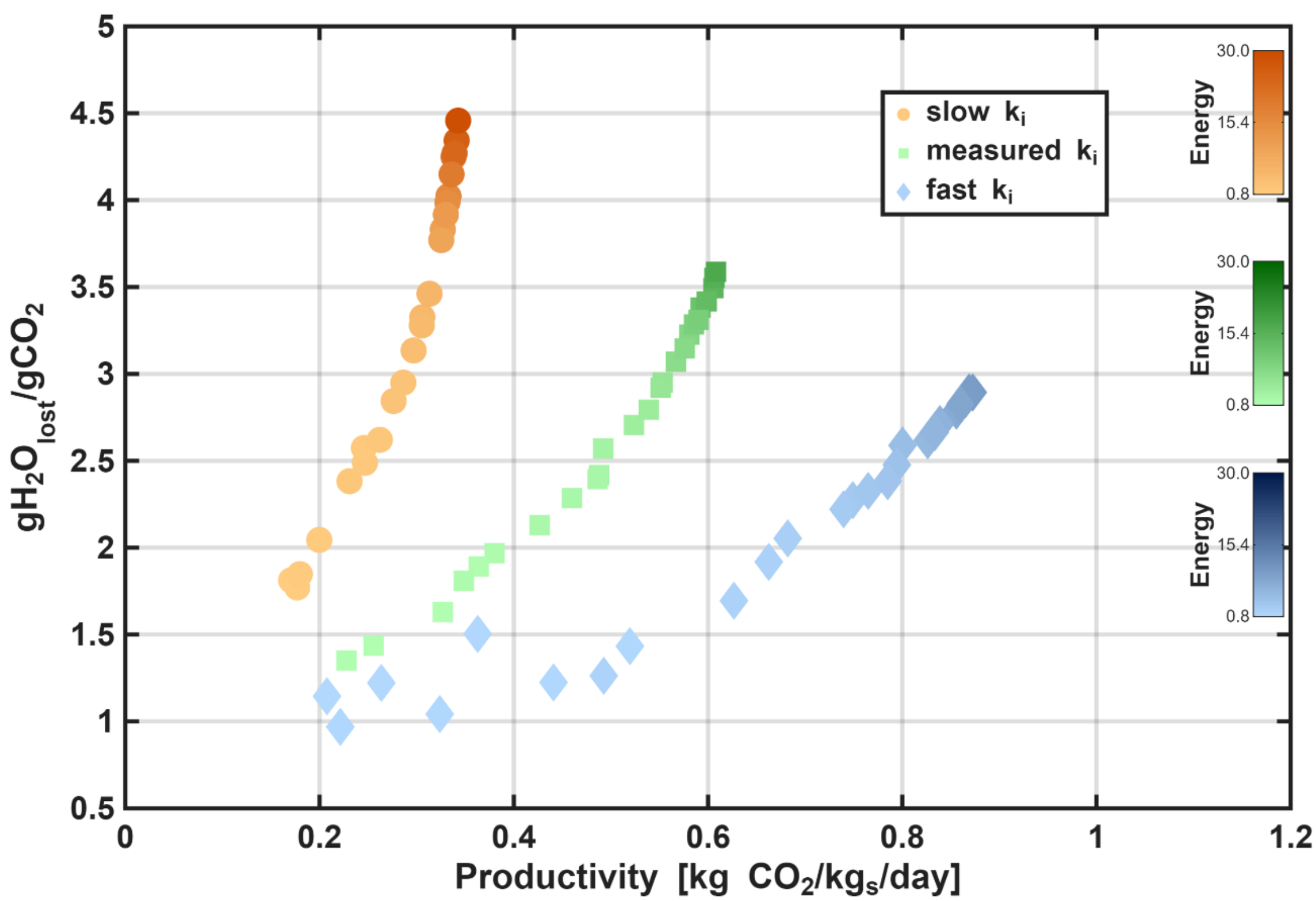

Fig. 12 12 Pareto fronts of water loss per $CO_2$ captured vs productivity for three different k scenarios—slow, measured (baseline) and fast at a fixed RH of 20%. The color bar represents the electrical energy required, ranging from 0.8-30 $[MJ \cdot kgCO_2{}^{-1}]$.

**Table 7 Varied kinetic sorption and desorption values for $CO_2$ and $H_2O$ presented as slow (0.5x baseline), measured (1x baseline) and fast (2x baseline).**

| k | $k_{CO_2}^{ads}$ $[s^{-1}]$ ($x10^{-4}$) | $k_{CO_2}^{des}$ $[s^{-1}]$ ($x10^{-4}$) | $k_{H_2O}^{ads}$ $[s^{-1}]$ ($x10^{-4}$) | $k_{H_2O}^{des}$ $[s^{-1}]$ ($x10^{-4}$) |
|---|---|---|---|---|
| Slow | 2.5 | 2.0 | 2.3 | 7.0 |
| Measured | 5.1 | 4.0 | 4.5 | 14 |
| Fast | 10 | 8.0 | 9.0 | 28 |

The plot in Fig. 12 clearly shows how faster kinetics drive the Pareto forward, resulting in greater productivity at lower water loss. The same trend is true for the energy penalty, though to a lesser extent. To contextualize these trends, **Table 8** lists the operating conditions corresponding to optimized points for minimum energy penalty, maximum productivity, and minimum water loss. Relative to the measured baseline, doubling the mass transfer coefficients increases maximum productivity by 43% (0.87 vs. 0.61 $kg\ CO_2 \cdot kg_s^{-1} \cdot d^{-1}$), decreases water loss by 19% (2.9 vs. 3.6 $kgH_2O \cdot kgCO_2{}^{-1}$), and reduces the energy penalty by 83% (9.3 vs. 17 $MJ \cdot kgCO_2{}^{-1}$). Conversely, halving the mass transfer coefficients decreases maximum productivity by 44%, increases water loss by 25%, and raises the energy penalty by 76%. These results confirm that the vapor and $CO_2$ mass transfer rates strongly govern overall VMS performance.

**Table 8 Optimized operating conditions at three kinetic levels - slow, measured (section 3.1) and fast at a fixed RH of 20% that correspond to minimum energy, maximum productivity, and minimum water loss. The modified kinetic rate constants for $CO_2$ and $H_2O$ are given in Table 7.**

| k | Objective | $t_{CS}$ [min] | $t_{VS}$ [min] | $t_{DES}$ [min] | $v_{AIR}$ [m/s] | $v_{VS}$ [m/s] | $E_T$ [MJ/ kg $CO_2$] | $Pr$ [kg$CO_2$ $kg_s^{-1}$ $day^{-1}$] | $H_2O$ Loss [g$H_2O$/ g$CO_2$] |
|---|---|---|---|---|---|---|---|---|---|
| slow | (Min $E_T$) | 145 | 146 | 6.2 | 0.15 | 0.22 | 0.9 | 0.17 | 1.8 |
| | (Max Pr) | 54 | 64 | 4.6 | 0.98 | 5.0 | 30 | 0.34 | 4.5 |
| | (Min $H_2O$ loss) | 145 | 146 | 6.3 | 0.25 | 0.22 | 0.9 | 0.18 | 1.8 |
| measured | (Min $E_T$) | 109 | 93 | 3.7 | 0.33 | 0.22 | 1.0 | 0.26 | 1.4 |
| | (Max Pr) | 39 | 51 | 3.9 | 0.95 | 5.0 | 17 | 0.61 | 3.6 |
| | (Min $H_2O$ loss) | 120 | 104 | 3.7 | 0.42 | 0.16 | 1.1 | 0.23 | 1.4 |
| fast | (Min $E_T$) | 50 | 162 | 4.5 | 0.40 | 0.11 | 0.9 | 0.26 | 1.2 |
| | (Max Pr) | 43 | 45 | 4.4 | 0.97 | 4.45 | 9.3 | 0.87 | 2.9 |
| | (Min $H_2O$ loss) | 53 | 162 | 5.0 | 0.5 | 0.05 | 1.1 | 0.22 | 1.0 |

Operationally, the highest productivity is achieved with short step durations in both the air and vapor stripping stages, coupled with faster water vapor velocities that promote rapid sorbent saturation and regeneration within short cycle times. In contrast, minimizing water loss favors longer cycle durations and slower vapor velocities. This approach reduces evaporative losses relative to $CO_2$ capture, but at the expense of both productivity due to diminishing sorption returns during extended air contact, and energy efficiency since prolonged air flow rates over the sorbent bed increases energetic load. Together, these results emphasize that VMS operation is bounded by a coupled trade-off between kinetics, water management, and energy input.

Opportunities to reduce electrical demand across vacuum-based DAC systems therefore center on minimizing pressure drop while preserving high mass-transfer rates. Passive air-contacting approaches that leverage ambient wind-driven advection can reduce blower energy but at the expense of sorbent utilization and system productivity.57 Alternatively, engineered sorbent architectures—such as sorbent-coated monolithic channels—offer a pathway to lower pressure drop without sacrificing mass transfer, analogous to catalytic converter designs used in automotive exhaust systems.58–60 In contrast, solvent-based DAC systems typically experience lower pressure drop in air contactors, but incur additional electrical energy demand associated with oxygen production for oxy-fuel combustion, which enables recovery of both combustion-derived $CO_2$ and $CO_2$ captured from air.

Opportunities to reduce electrical demand across vacuum-based DAC systems therefore center on minimizing pressure drop while preserving high mass-transfer rates. Passive air-contacting approaches that leverage ambient wind-driven advection can reduce blower energy but at the expense of sorbent utilization and system productivity.57 Alternatively, engineered sorbent

architectures—such as sorbent-coated monolithic channels—offer a pathway to lower pressure drop without sacrificing mass transfer, analogous to catalytic converter designs used in automotive exhaust systems.58–60 In contrast, solvent-based DAC systems typically experience lower pressure drop in air contactors, but incur additional electrical energy demand associated with oxygen production for oxy-fuel combustion, which enables recovery of both combustion-derived $CO_2$ and $CO_2$ captured from air.

## 3.5 Benchmarking VMS against leading DAC technology

To place these operational trade-offs in context, we benchmark established DAC technologies reported in the literature to the optimized VMS process (20% RH, baseline kinetics from section 3.3). Table 9 compares productivity, thermal and electrical energy demand, and water management characteristics of VMS with leading sorbent and solvent-based capture technologies. Comparisons are made to simulated optimizations of TVSA[31] and steam-assisted TVSA systems[23,29], as well as pilot plant data.[16,30] In s-TVSA, steam is used both to heat the supported amine sorbent for thermal desorption (95-120 °C) and to advect the released $CO_2$ out of the sorbent bed, thereby increasing the driving force for desorption. This role is analogous to the low-temperature water vapor used in VMS where moisture driven desorption replaces the thermal desorption required for amine sorbents. We further compare VMS to a commercially demonstrated liquid-absorption system, as presented in the first detailed technoeconomic analysis of scaled DAC.[16] In that case the 900 °C thermal load (supplied with natural gas oxy-combustion) and electrical loads (oxygen separation, compression, pumping) are considered separately. To maintain a consistent basis across studies, we subtract the electrical energy for $CO_2$ compression above 0.1MPa, matching compression levels considered here (Equation 27) and in the amine-based comparisons.

The top row in Table 9 compares $CO_2$ productivity across the various DAC technologies. The VMS value in bold represents the Pareto inflection point, case II in Fig. 8, while the range of values in parentheses represents the Pareto-optimal solutions from lowest-productivity, lowest-energy consumption regime (0.23 $\mathrm{kg\,CO_2 \cdot kg_s^{-1} \cdot d^{-1}}$, 1.1 $\mathrm{MJ \cdot kgCO_2}^{-1}$ , and 1.4 $\mathrm{kgH_2O \cdot kgCO_2}^{-1}$) to the highest-productivity operating point that incurs the greater water loss and energy penalty (0.61 $\mathrm{kg\,CO_2 \cdot kg_s^{-1} \cdot d^{-1}}$, 15 $\mathrm{MJ \cdot kgCO_2}^{-1}$ , and 3.5 $\mathrm{kgH_2O \cdot kgCO_2}^{-1}$). This range overlaps with the three steam-assisted TVSA cases, including pilot plant results obtained using a macroporous amine functionalized resin (Lewatit VP OC 1065), which shares structural features with the moisture-responsive IRA900.[39] Without a steam sweep, simple TVSA relies solely on pressure rise and vacuum pumping for $CO_2$ desorption and transport, and under all scenarios VMS and s-TVSA outperforms TVSA in productivity. The associated range of sorbent working capacity and cycle duration, which together determine productivity, are shown at the bottom of Table 9. VMS and TVSA-based approaches generally cycle the sorbent between 0.6-1.2 mmol/g, with smaller working capacities resulting from faster cycling. The pilot plant s-TVSA process reports a cycle duration of 200 minutes, which falls close to the higher range of the VMS-optimized cycle length range.

Comparison with the liquid absorption system is more complicated as this technology operates continuously and couples air contacting solvent absorption to a calcium oxide calcination loop (i.e. $Ca(OH)_2/CaCO_3/CaO$ in a pellet reactor, calciner and slaker). Productivity for this process is calculated on the basis of the regenerated solid product leaving the 900 °C calciner and does not consider the sizing or mass throughput of the upstream liquid air contactor. This continuous solids-circulation basis explains the order-of-magnitude higher productivity calculated for the liquid-absorption relative to solid-sorbent DAC.

**Table 9. Performance comparison of the optimized VMS process (using 20% RH air and measured kinetics) against established DAC technologies. For VMS data, bold values represent the Pareto front inflection point (Case ii, Fig. 8, Table 5), while parentheses indicate the full range of non-weighted optimization results.**

| **Process** | **VMS** | **s- TVSA** | | | **TVSA** | **Liquid Absorption** |
|---|---|---|---|---|---|---|
| **Material** | **IRA 900 (simulation)** | **APDES (simulation)** | **APDES NFC (simulation)** | **Lewatit VP OC 1065 (pilot plant)** | **Lewatit VP OC 1065 (simulation)** | **KOH (pilot plant)** |
| Productivity $\left[kgCO_2 \cdot kg_s^{-1} \cdot day^{-1}\right]$ | **0.5** (0.23 – 0.61) | 0.13 – 0.23† | 0.22 | 0.29 | 0.06 – 0.1‡ | 15§ |
| Thermal energy $\left[MJ \cdot kgCO_2{}^{-1}\right]$ | **0** | 7 – 48 | 18-40 | 12 | 8.8-13 | 5.3 |
| Electrical energy¶ $\left[MJ \cdot kgCO_2{}^{-1}\right]$ | **2.5** (1 – 15) | 1.0 | 0.34 | 2.7 | 0.8-1.8 | 0.84 |
| Water loss $\left[kgH_2O \cdot kgCO_2{}^{-1}\right]$ | **2.4** (1.4 – 3.5) | 0 | 0 | 0 | 0 | 7.5‖ |
| Water processed $\left[kgH_2O \cdot kgCO_2{}^{-1}\right]$ | **133** (34 – 360) | 95-169 | 22-102 | 1.7 | nm | 290# |
| Regeneration temperature [°C] | **40** (30-50) | 120 | 95 | 110 | 100 | 900 |
| Working capacity $[mol\ kg^{-1}]$ | **0.6** (0.58-1.0) | 0.9 – 1.2 | 1.8 | 0.91 | 0.7-1.1 | na |
| Cycle duration (min) | **102** (94 – 207) | 420 – 660 | 742 | 200 | 515 | continuous |
| Reference | current work | 23 | 29 | 30 | 31 | 16 |

na = not applicable. nm = not mentioned †Range based on vacuum pressure of desorption (5-30 kPa). ‡ Range based on three co-adsorption isotherm models. §Productivity based on rate of cycled calcium oxide/potassium carbonate from the calciner relative to the rate of $CO_2$ captured from air. ¶Electrical energy includes $CO_2$ compression up to 0.1 MPa. ‖Water loss from 2M KOH solution contacting 20 °C air at 35% RH. #Water processed based on 2M KOH circulation.

Beyond productivity, the thermal energy comparison in Table 9 highlights the key distinguishing feature of the VMS. Unlike leading DAC technologies, VMS requires no external thermal input. The moderate 12-27 K temperature rise observed during the vapor stripping step (Fig. 7(d)) is entirely autothermal and originates from the predominate water-vapor sorption exotherm, and is sensitive to the vapor flow rate (Fig. S2(f)). As a result, VMS avoids dependence on steam generation or external heaters – requirements that dominate the energetic loads of other systems and impacts the capital expense of the system. For example, high temperature calcination incurs thermal demands of approximately 5.3 $\mathrm{MJ \cdot kgCO_2}^{-1}$ and additional electricity demands to separate oxygen to ensure the $CO_2$ generated from natural gas combustion is also captured. Moderate temperature thermal amine regeneration requires anywhere from 7-48 $\mathrm{MJ \cdot kgCO_2}^{-1}$. In steam-assisted TVSA, the thermal energy span is driven by the volumetric flow of steam required to accelerate $CO_2$ desorption. In both TVSA approaches, the thermal energy is also affected by the need to desorb any co-adsorbed water vapor from air as sorbents enter the $CO_2$ sorption stage in a dry state. A further constraint of amine-based sorbents is their limited thermal oxidative stability, which limits regeneration temperatures near ~ 95°C and requires significant sorbent makeup, leading to an ongoing operating expense of the process.[20,21] Because VMS operates at low temperatures, thermal degradation is not a concern; however, the ambient stability of different moisture-swing sorbents remains an important area for continued study.

Across all DAC technologies, electrical energy is required for air handling through sorbent bends or solvent contactors and for compressing captured $CO_2$ to pressures suitable for downstream use (e.g. pipeline transport for geologic sequestration or ambient pressure utilization). Because treated air volumes are large, pressure drop across air-contacting structures is a large contributor to electrical demand. As a result, nearly all DAC configurations – including VMS – operate at broadly similar air velocities, typically limited to below ~1 $\mathrm{m \cdot s^{-1}}$ to balance productivity against blower power. Lower air velocities reduce electricity consumption but extend sorption times and suppress productivity, whereas operation near this practical velocity limit shortens cycle times and increases productivity at the cost of higher electrical input. Electrical energy requirements associated with air contacting can be reduced by employing passive designs that leverage ambient wind-driven advection; however, such approaches come at the expense of sorbent utilization and overall system productivity.[57] Alternatively, sorbent bed architectures can be made to lower pressure drop while maintaining high mass transfer, often achieved through the use of sorbent-coated monolithic channels,[58–60] analogous to catalytic converter configurations in automotive exhaust systems. In contrast, solvent-based DAC alleviate pressure drop limitations in the air contactors but require additional demand to generate oxygen for oxy-fuel natural gas combustion, enabling recovery of both the combustion-derived $CO_2$ with the $CO_2$ captured from air.

Differences in electrical demand across sorbent-based DAC systems are largely governed by regeneration dynamics. In both VMS and steam-assisted TVSA, additional electrical energy is required to transport water vapor or steam through the sorbent bed during regeneration. In VMS, the vapor-stripping flow rate can dominate the energy-productivity tradeoff (Fig. 10), where higher vapor velocities accelerate regeneration and enable high productivity but drive the upper range of electrical energy demand and increase water loss per unit $CO_2$ captured. At moderate vapor velocities, however, the electrical energy requirement of VMS (~2.5 $\mathrm{MJ \cdot kgCO_2}^{-1}$) aligns closely with those of optimized TVSA systems, while retaining the defining advantage of VMS in eliminating any supplemental thermal energy input.

When DAC electrical requirements are considered alongside thermal loads, VMS exhibits markedly lower total energy consumption relative to all other leading technologies considered. Consequently, any direct $CO_2$ emissions associated with energy generation are proportionally smaller for VMS, thereby improving its net $CO_2$ removal efficiency and enhancing its effective productivity compared to systems with overall higher energy intensity.

While the absence of external thermal energy confers a distinct advantage to VMS, this benefit is accompanied by a trade-off in water use. Because VMS relies on moisture-driven regeneration rather than thermal inputs, the energy that would otherwise be supplied as heat is instead provided by the latent heat of water evaporation that drives cyclic $CO_2$ binding (eqn 1). Consequently, evaporative water loss from the sorbent to air during $CO_2$ sorption represents a real and quantifiable resource cost. As shown previously in Fig. 11, the 1.4-3.5 $\mathrm{kgH_2O \cdot kgCO_2}^{-1}$ arises from a productivity trade-off. Higher productivity is achieved at shorter cycle lengths and faster vapor speed. The faster water vapor flow enables faster and deeper $H_2O$ sorption and $CO_2$ desorption across a cycle (Figs. S2(e) and S2(d)) which elevates the amount of water loss during the $CO_2$ sorption step.

Beyond the water lost, even more water is processed in the vapor stripping step. The wide range of total water processed, 34 - 360 $\mathrm{kgH_2O \cdot kgCO_2}^{-1}$ , reflects the span of optimized vapor stripping velocities and durations. Downstream of the sorbent bed, the heat released during compression of the vapor-gas mixture will be transferred to the cold-water reservoir in a condenser, eliminating the need for supplemental cooling to condense the vapor or to reheat the feed-water supply cooled from evaporation. The recovered liquid water can be recycled within the process or depending on regional water-supply and demand conditions, treated as an upgraded byproduct. Importantly, the vacuum-driven evaporation central to VMS enables the use of brackish or low-quality water rather than freshwater. Because evaporation inherently rejects most dissolved salts and impurities, the process avoids direct competition with freshwater resources – an increasingly critical constraint in many arid regions. As a result, VMS water demand may not pose a major limitation, provided saline or brackish water is locally available, and appropriate design measures prevent salt deposition or volatile organic contamination within the sorbent bed and ancillary equipment.

Water management considerations also arise in s-TVSA systems, where water is introduced through the steam sweep. In the pilot-plant study, low steam flow rates resulted in modest water processing (1.7 $\mathrm{kgH_2O \cdot kgCO_2}^{-1}$), while higher sweep rates approaching those of VSA lead to substantially larger water throughput. Even TVSA systems without steam assistance process water through humidity uptake from ambient air, followed by thermal desorption during regeneration.

Although this quantity is proportional to ambient relative humidity and $CO_2$:$H_2O$ selectivity of the amine sorbent, it remains far lower than the water demands associated with the VMS or s-TVSA. In contrast, the liquid solvent system circulates an aqueous 2M KOH solvent, resulting in water processing of approximately 290 $\mathrm{kgH_2O \cdot kgCO_2}^{-1}$. Thus, the VMS process is broadly comparable to solvent-based DAC in terms of both water loss and water throughput. However, VMS technology will benefit from more arid to moderately humid climates, resulting in greater productivity and energy-efficiency, albeit at the cost of water loss (Fig 11), while solvent-based separations will benefit from more humid climates to limit solvent water loss. This distinction highlights how regional climate and resources (e.g. water and power) shapes the suitability of different carbon removal options.

Beyond quantifying the water-energy trade-off, this work introduces a new VMS process that broadens the design space for DAC through a low-temperature, moisture-driven regeneration pathway. The coupled $CO_2$/$H_2O$ kinetics experimentally established here, together with newly established MS isotherms,[43] provide the quantitative foundation data needed for detailed technoeconomic (TEA) and life cycle analyses (LCA). These data enable assessments that incorporate $CO_2$ end-use requirements, the $CO_2$ intensity of regional electrical grids, equipment-scale considerations, and local water resource constraints. Such economic and systems-level comparisons will be essential for identifying where, and what type of, DAC deployment is most advantageous – ideally in proximity to safe, permanent sequestration opportunities.

Looking forward, a comprehensive pilot-plant study of the VMS process together with an assessment of sorbent durability over $>10^5$ cycles will be essential for evaluating sorbent-related operating expenditures and validate process kinetics across a range of near ambient temperatures. Emerging low-cost activated-carbon impregnated carbonate salt sorbents that exhibit moisture-swing behavior offer additional opportunities to tailor material properties and optimize performance under different water-loading regimes.[41,61,62] Finally, integrating waste heat into the water reservoir or leveraging geothermally warm saline groundwater, such as that found in the Four Corners region of the U.S. Southwest,[63] may further reduce energy demands by increasing saturation pressure and potentially accelerating (de)sorption kinetics. Such strategies highlight opportunities for process intensification and ultimately lowering the cost of negative $CO_2$ emissions.

# 4. Conclusion

This work establishes the Vacuum Moisture Swing (VMS) as a promising low energy pathway for DAC, demonstrated here using a bicarbonate-exchanged anion exchange resin, IRA 900. By integrating experimentally validated $CO_2$/$H_2O$ sorption kinetics with a coupled mass-energy transport model, we show that VMS leverages adequate $CO_2$ sorption and water-vapor-driven regeneration to achieve competitive productivity (up to 0.61 kg $\mathrm{CO_2 \cdot kg_s^{-1} \cdot d^{-1}}$) without external thermal input. Multi-objective optimization reveals clear trade-offs between energy use, productivity and water loss, with air velocity governing electrical demand and vapor velocity controlling both sorbent utilization and evaporative losses. Parameterization across ambient relative humidity and kinetic regimes further demonstrates that VMS performance can be tuned to local ambient conditions and potentially improved through targeted enhancements to sorbent mass-transfer rates.

Benchmarking against leading DAC technologies highlights the distinct operating regime of the VMS: it avoids the substantial thermal penalties of TVSA and solvent based systems while matching or exceeding sorbent-based productivity at regeneration temperature below 40 $^oC$. Although VMS processes large quantities of water, most is recovered, resulting in net water losses of only 1.4-3.5 $\mathrm{kgH_2O \cdot kgCO_2}^{-1}$, comparable to solvent-based DAC. Further, the ability to utilize saline water minimizes competition with freshwater resources. These attributes suggest that VMS is particularly suited for deployment in arid regions where low humidity improves productivity and energy efficiency.

Together, these results broaden the design landscape for DAC by establishing a moisture-swing pathway with inherently low thermal requirements and regionally adaptable performance. The experimentally derived kinetics together with newly reported MS isotherms incorporated into this model provide the foundation for rigorous TEA and LCA assessments that incorporate local water availability, grid carbon intensity, and proximity to permanent storage. Advancing VMS toward deployment will require pilot-scale demonstrations, long-term sorbent durability testing, and continued material innovation – including the consideration of carbonate-impregnated porous sorbents – to further reduce cost, water and energy demand. Overall, VMS offers a compelling direction for scalable, low-energy $CO_2$ removal and expands the suite of DAC technologies to meet diverse climate and resource contexts.

## NOMENCLATURE

| | |
|---|---|
| $A$ | cross-sectional area of the column $[\mathrm{m}^2]$ |
| $c$ | gas phase concentration $[\mathrm{mol\ m}^{-3}]$ |
| $C$ | GAB model temperature dependency [-] |
| $C_{p,a}$ | specific heat capacity of the sorbed phase $[\mathrm{J\ mol}^{-1}\ \mathrm{K}^{-1}]$ |
| $C_{p,g}$ | specific heat capacity of the gas phase $[\mathrm{J\ mol}^{-1}\ \mathrm{K}^{-1}]$ |
| $C_{p,s}$ | specific heat capacity of the sorbent $[\mathrm{J\ kg}^{-1}\ \mathrm{K}^{-1}]$ |
| $D_L$ | axial dispersion $[\mathrm{m}^2\ \mathrm{s}^{-1}]$ |
| $D_m$ | molecular diffusivity $[\mathrm{m}^2\ \mathrm{s}^{-1}]$ |
| $d_p$ | particle diameter [m] |
| $E_{mo}$ | monolayer heats of sorption $[\mathrm{kJ\ mol}^{-1}]$ |
| $E_{ml}$ | multilayer heats of sorption $[\mathrm{kJ\ mol}^{-1}]$ |
| En | electrical energy for a given step [J] |
| $\mathrm{E_T}$ | total electrical energy for a cycle [J] |
| $H$ | enthalpy $[\mathrm{J\ mol}^{-1}]$ |
| $IEC$ | Ion exchange capacity $[\mathrm{mol\ kg}^{-1}]$ |
| $k_i$ | mass transfer coefficient for species $i$ $[\mathrm{s}^{-1}]$ |
| $k_i^{ads}$ | sorption kinetics $[\mathrm{s}^{-1}]$ |
| $k_i^{des}$ | desorption kinetics $[\mathrm{s}^{-1}]$ |
| $K_{ads}$ | GAB model temperature dependency [-] |
| $K_{MS}$ | $CO_2$ sorption isotherm equilibrium constant |
| $K_z$ | effective gas thermal conductivity $[\mathrm{J\ m}^{-1}\ \mathrm{K}^{-1}\ \mathrm{s}^{-1}]$ |
| $L$ | column length [m] |
| $n_i$ | moles of species $i$ [mol] |
| $n$ | $CO_2$ isotherm power law [-] |
| $MW_i$ | molecular weight of species $i$ $[\mathrm{kg}\cdot\mathrm{mol}^{-1}]$ |
| $N$ | number of discretization volumes |
| $P$ | high pressure [Pa] |
| $Pr$ | productivity $[\mathrm{kg\ CO_2}\cdot\mathrm{kg_s}^{-1}\cdot\mathrm{d}^{-1}]$ |
| $P\mathrm{CO_2}$ | Partial pressure of $CO_2$ [Pa] |
| $q$ | concentration in solid phase $[\mathrm{mmol\ g}^{-1}]$ |
| $q^*$ | equilibrium loading in solid phase $[\mathrm{mmol\ g}^{-1}]$ |
| $q_m$ | monolayer heats of sorption $[\mathrm{mol\ kg}^{-1}]$ |
| $q\mathrm{CO_2}$ | $CO_2$ loading $[\mathrm{mol\ m}^{-3}]$ |
| $q\mathrm{H_2O}$ | water loading $[\mathrm{mol\ m}^{-3}]$ |
| $r_{in}$ | column inner radius [m] |
| $r_{out}$ | column outer radius [m] |
| $r_p$ | particle radius [m] |
| $R$ | universal gas constant $[\mathrm{Pa\ m}^3\ \mathrm{mol}^{-1}\ \mathrm{K}^{-1}]$ |
| $RH$ | relative humidity [%] |
| $t$ | time [s] |
| $T$ | temperature [K] |
| $v$ | velocity $[\mathrm{m\ s}^{-1}]$ |
| z | Bed coordinate [m] |

**Greek Symbols**

| | |
|---|---|
| $\varepsilon_b$ | bed voidage [-] |
| $\varepsilon_p$ | particle voidage [-] |
| $\eta$ | turbomachinery efficiency [-] |
| $\gamma$ | adiabatic constant [-] |
| $\lambda$ | pressure rate constant [-] |
| $\mu$ | fluid viscosity [kg m$^{-1}$ s$^{-1}$] |
| $\rho_g$ | density of gas phase [kg m$^{-3}$] |
| $\rho_s$ | density of sorbent [kg m$^{-3}$] |
| $\tau$ | tortuosity [-] |

**Subscript**

| | |
|---|---|
| $CS$ | $CO_2$ sorption |
| $PZ$ | pressurization |
| $EV$ | evacuation |
| $VS$ | vapor stripping |
| $DES$ | desorption |
| *H* | high |
| *I* | intermediate |
| *L* | low |
| *feed* | feed gas |
| *in* | stream entering |
| *out* | stream leaving |
| *AIR* | inlet air |
| $i$ | index of component |
| *sat* | Saturation |
| *int* | Interstitial |
| *sup* | superficial |

# Corresponding Author

*Jennifer Wade. Steve Sanghi College of Engineering, Department of Mechanical Engineering,

Northern Arizona University, Flagstaff, Arizona 86011, USA

Email: Jennifer.Wade@nau.edu

## Author Contributions

**S.S.** led the data acquisition, software development, data analysis, visualization, writing – original draft, review & editing

**S.B.** prepared samples and established experimental method development.

**G.B.** led and evaluated the kinetic investigation, writing – original draft.

**P.S.** project conceptualization, methodology, writing – original draft.

**J.L.W.*** project conceptualization, funding acquisition, visualization, supervision, writing – original draft, review & editing.

## Declaration of interest

The authors declare no competing interests.

## Acknowledgements

This material is based upon work primarily supported by the National Science Foundation under Grant No. 2219247 (SS, SB and GB). Further, SB, GB and JLW were partially supported by the U.S. Department of Energy, Office of Science, Office of Basic Energy Sciences under Award Number DE-SC0023343. The authors further acknowledge Arizona State University's Center for Negative Carbon Emissions for their collaborative guidance and critical discussions in shaping this research over the past several years

# References

1 Carbon neutrality by 2050: the world's most urgent mission | United Nations Secretary-General, https://www.un.org/sg/en/content/sg/articles/2020-12-11/carbon-neutrality-2050-the-world%E2%80%99s-most-urgent-mission, (accessed 13 January 2024).

2 L. Chen, G. Msigwa, M. Yang, A. I. Osman, S. Fawzy, D. W. Rooney and P.-S. Yap, *Environ. Chem. Lett.*, 2022, **20**, 2277–2310.

3 S. Fuss, J. G. Canadell, P. Ciais, R. B. Jackson, C. D. Jones, A. Lyngfelt, G. P. Peters and D. P. Van Vuuren, *One Earth*, 2020, **3**, 145–149.

4 J. C. Minx, W. F. Lamb, M. W. Callaghan, S. Fuss, J. Hilaire, F. Creutzig, T. Amann, T. Beringer, W. de Oliveira Garcia, J. Hartmann, T. Khanna, D. Lenzi, G. Luderer, G. F. Nemet, J. Rogelj, P. Smith, J. L. Vicente Vicente, J. Wilcox and M. del Mar Zamora Dominguez, *Environ. Res. Lett.*, 2018, **13**, 063001.

5 N. A. of S. Medicine Engineering, and, D. on E. and L. Studies, O. S. Board, B. on C. S. and Technology, B. on E. S. and Resources, B. on A. and N. Resources, B. on E. and E. Systems, B. on A. S. and Climate and C. on D. a R. A. for C. D. R. and R. Sequestration, *Negative Emissions Technologies and Reliable Sequestration: A Research Agenda*, National Academies Press, 2019.

6 R. S. Haszeldine, S. Flude, G. Johnson and V. Scott, *Philos. Trans. R. Soc. Math. Phys. Eng. Sci.*, 2018, **376**, 20160447.

7 S. Fawzy, A. I. Osman, J. Doran and D. W. Rooney, *Environ. Chem. Lett.*, 2020, **18**, 2069–2094.

8 The emerging global crisis of land use | 03 Land and climate pressures, https://www.chathamhouse.org/2023/11/emerging-global-crisis-land-use/03-land-and-climate-pressures, (accessed 6 January 2025).

9 L. Rosa, D. L. Sanchez, G. Realmonte, D. Baldocchi and P. D'Odorico, *Renew. Sustain. Energy Rev.*, 2021, **138**, 110511.

10 M. Erans, E. S. Sanz-Pérez, D. P. Hanak, Z. Clulow, D. M. Reiner and G. A. Mutch, *Energy Environ. Sci.*, 2022, **15**, 1360–1405.

11 A. Sodiq, Y. Abdullatif, B. Aissa, A. Ostovar, N. Nassar, M. El-Naas and A. Amhamed, *Environ. Technol. Innov.*, 2023, **29**, 102991.

12 H. Xu, L. Yu, C. Chong and F. Wang, *Energy Convers. Manag.*, 2024, **322**, 119119.

13 R. Custelcean, N. J. Williams, K. A. Garrabrant, P. Agullo, F. M. Brethomé, H. J. Martin and M. K. Kidder, *Ind. Eng. Chem. Res.*, 2019, **58**, 23338–23346.

14 J. Young, N. McQueen, C. Charalambous, S. Foteinis, O. Hawrot, M. Ojeda, H. Pilorgé, J. Andresen, P. Psarras, P. Renforth, S. Garcia and M. van der Spek, *One Earth*, 2023, **6**, 899–917.

15 P. Priyadarshini, G. Rim, C. Rosu, M. Song and C. W. Jones, *ACS Environ. Au*, 2023, **3**, 295–307.

16 D. W. Keith, G. Holmes, D. St. Angelo and K. Heidel, *Joule*, 2018, **2**, 1573–1594.

17 K. An, K. Li, C.-M. Yang, J. Brechtl and K. Nawaz, *J. CO2 Util.*, 2023, **76**, 102587.

18 R. Custelcean, *Annu. Rev. Chem. Biomol. Eng.*, 2022, **13**, 217–234.

19 R. Custelcean, K. A. Garrabrant, P. Agullo and N. J. Williams, *Cell Rep. Phys. Sci.*, 2021, **2**, 100385.

20 Q. T. Vu, H. Yamada and K. Yogo, *Ind. Eng. Chem. Res.*, 2021, **60**, 4942–4950.

21 H. E. Holmes, S. Banerjee, A. Vallace, R. P. Lively, C. W. Jones and M. J. Realff, *Energy Environ. Sci.*, 2024, **17**, 4544–4559.

22 W. K. Shi, X. J. Zhang, X. Liu, S. Wei, X. Shi, C. Wu and L. Jiang, *J. Clean. Prod.*, 2023, **414**, 137731.
23 B. M. Balasubramaniam, P.-T. Thierry, S. Lethier, V. Pugnet, P. Llewellyn and A. Rajendran, *Chem. Eng. J.*, 2024, **485**, 149568.
24 T. Deschamps, M. Kanniche, L. Grandjean and O. Authier, *Clean Technol.*, 2022, **4**, 258–275.
25 W. Liu, Y. Ji, R. Q. Wang, X. J. Zhang and L. Jiang, *Appl. Energy*, 2023, **335**, 120757.
26 F. Donat and C. R. Müller, *Curr. Opin. Green Sustain. Chem.*, 2022, **36**, 100645.
27 J. Hu, Y. Jiang, Q. Gao, Y. Zhao, S. Dai, X. Li and W. Wei, *Chem. Eng. J.*, 2025, **505**, 159237.
28 H. Zentou, B. Hoque, M. A. Abdalla, A. F. Saber, O. Y. Abdelaziz, M. Aliyu, A. M. Alkhedhair, A. J. Alabduly and M. M. Abdelnaby, *Carbon Capture Sci. Technol.*, 2025, **15**, 100386.
29 V. Stampi-Bombelli, M. van der Spek and M. Mazzotti, *Adsorption*, 2020, **26**, 1183–1197.
30 H. M. Schellevis and D. W. F. Brilman, *React. Chem. Eng.*, 2024, **9**, 910–924.
31 J. Young, E. García-Díez, S. Garcia and M. van der Spek, *Energy Environ. Sci.*, 2021, **14**, 5377–5394.
32 J. Young, F. Mcilwaine, B. Smit, S. Garcia and M. van der Spek, *Chem. Eng. J.*, 2023, **456**, 141035.
33 A. Sinha, L. A. Darunte, C. W. Jones, M. J. Realff and Y. Kawajiri, *Ind. Eng. Chem. Res.*, 2017, **56**, 750–764.
34 A. Sodiq, Y. Abdullatif, B. Aissa, A. Ostovar, N. Nassar, M. El-Naas and A. Amhamed, *Environ. Technol. Innov.*, 2023, **29**, 102991.
35 S. Kim, A. Sarswat, S. Cho, M. J. Realff, D. S. Sholl and R. Lively, *Social Science Research Network*, 2024, preprint, Social Science Research Network:4818768, DOI: 10.2139/ssrn.4818768.
36 Y. Wang, J. Kim, J. Marreiros, N. Rangnekar, Y. Yuan, J. R. Johnson, B. A. McCool, M. J. Realff and R. P. Lively, *ACS Sustain. Chem. Eng.*, 2025, **13**, 6554–6564.
37 Y. Zhu, A. Booth and K. B. Hatzell, *Environ. Sci. Technol. Lett.*, 2024, **11**, 89–94.
38 J. Hegarty, B. Shindel, D. Sukhareva, M. L. Barsoum, O. K. Farha and V. Dravid, *Environ. Sci. Technol.*, 2023, **57**, 21080–21091.
39 G. Najaf Tomaraei, S. Binney, R. Stratton, H. Zhuang and J. L. Wade, *ACS Appl. Mater. Interfaces*, 2025, **17**, 53547–53562.
40 I. Nicotera, A. Enotiadis and C. Simari, *Small*, 2024, **20**, 2401303.
41 J. Mendez Lozoya, E. Solis Mata, J. J. Velazquez Salazar, A. Hernandez Cedillo, M. J. Yacaman and J. L. Wade, *ACS Appl. Eng. Mater.*, 2025, acsaenm.5c00716.
42 DuPont™ AmberLite™ IRA900 Cl, https://www.dupont.com/products/amberliteira900cl.html, (accessed 16 November 2025).
43 H. Lopez-Marques, J. L. Wade, K. K. Reimund, S. K. Sinyangwe, S. Guzzo, L. A. Smith, A. Chamoun-Farah, H. Oh, W. Wang, P. R. Gupta, C. B. Mullins, M. Kumar and B. D. Freeman, *Environ. Sci. Technol.*, 2026, **60**, 1781–1790.
44 J. A. Wurzbacher, C. Gebald, S. Brunner and A. Steinfeld, *Chem. Eng. J.*, 2016, **283**, 1329–1338.
45 C. Gebald, J. A. Wurzbacher, A. Borgschulte, T. Zimmermann and A. Steinfeld, *Environ. Sci. Technol.*, 2014, **48**, 2497–2504.
46 L. Amiri, S. A. Ghoreishi-Madiseh, F. P. Hassani and A. P. Sasmito, *Powder Technol.*, 2019, **356**, 310–324.

47 Principles of Adsorption and Adsorption Processes Ruthven | PDF, https://www.scribd.com/document/538675127/Principles-of-Adsorption-and-Adsorption-Processes-Ruthven, (accessed 22 December 2023).
48 R. Haghpanah, A. Majumder, R. Nilam, A. Rajendran, S. Farooq, I. A. Karimi and M. Amanullah, *Ind. Eng. Chem. Res.*, 2013, **52**, 4249–4265.
49 G.-S. Jiang and C.-W. Shu, *J. Comput. Phys.*, 1996, **126**, 202–228.
50 X.-D. Liu, S. Osher and T. Chan, *J. Comput. Phys.*, 1994, **115**, 200–212.
51 PEESEgroup/PSA PEESE 2025.
52 K. T. Leperi, R. Q. Snurr and F. You, *Ind. Eng. Chem. Res.*, 2016, **55**, 3338–3350.
53 H2O and CO2 Sorption in Ion-Exchange Sorbents: Distinct Interactions in Amine Versus Quaternary Ammonium Materials | ACS Applied Materials & Interfaces, https://pubs.acs.org/doi/10.1021/acsami.5c12939, (accessed 27 October 2025).
54 S. Sarbaz, Z. X. Liu, H. Feigenbaum, S. Bayati, W. Wang, J. Wade, H. Mithaiwala and M. D. Green, *Polym. Test.*, 2025, **153**, 109024.
55 M. G. Marino, J. P. Melchior, A. Wohlfarth and K. D. Kreuer, *J. Membr. Sci.*, 2014, **464**, 61–71.
56 Seawater - Properties, https://www.engineeringtoolbox.com/sea-water-properties-d_840.html, (accessed 27 November 2025).
57 M. A. Talha, J. Cirucci and K. Lackner, *Research Square*, 2025, preprint, DOI: 10.21203/rs.3.rs-7782108/v1.
58 Y. J. Min, J. Kim, C. W. Jones and M. J. Realff, *ACS Sustain. Chem. Eng.*, 2024, **12**, 18522–18536.
59 P. De Joannis, C. Castel, M. Kanniche, E. Favre and O. Authier, *Carbon Capture Sci. Technol.*, 2025, **17**, 100518.
60 E. Tegeler, Y. Cui, M. Masoudi, A. M. Bahmanpour, T. Colbert, J. Hensel and V. Balakotaiah, *Chem. Eng. Sci.*, 2023, **281**, 119107.
61 S. Zheng, X. Cheng, W. Zhou, T. Wang, L. Zhu, H. Xiao and X. Chen, *Asia-Pac. J. Chem. Eng.*, 2024, **19**, e3058.
62 B. Shindel, J. Hegarty, J. D. Estradioto, M. L. Barsoum, M. Yang, O. K. Farha and V. P. Dravid, *Environ. Sci. Technol.*, 2025, **59**, 8495–8505.
63 J. S. Stanton, D. W. Anning, C. J. Brown, R. B. Moore, V. L. McGuire, A. C. Harris, K. F. Dennehy, P. B. McMahon and J. K. Bohlke, *Brackish groundwater in the United States: x*, U.S. Geological Survey, 2017.